# Identifying structure-absorption relationships and predicting absorption strength of non-fullerene acceptors for organic photovoltaics

*Jun Yan,[a,#] Xabier Rodríguez-Martínez,*[b,c,#] Drew Pearce,[a] Hana Douglas,[a] Danai Bili,[a] Mohammed Azzouzi,[a] Flurin Eisner,[a] Alise Virbule,[a] Elham Rezasoltani,[a] Valentina Belova,[c] Bernhard Dörling,[c] Sheridan Few,[a,f] Anna A. Szumska,[a] Xueyan Hou,[a] Guichuan Zhang,[d] Hin-Lap Yip,[d,e] Mariano Campoy-Quiles *[c] and Jenny Nelson*[a]*

[#] J.Y. and X.R.-M. contributed equally to this work.

[a] Department of Physics, Imperial College London, SW7 2AZ, London, United Kingdom
Email: jenny.nelson@imperial.ac.uk

[b] Electronic and Photonic Materials (EFM), Department of Physics, Chemistry and Biology (IFM), Linköping University, Linköping, SE 581 83 Sweden
Email: xabier.rodriguez.martinez@liu.se

[c] Instituto de Ciencia de Materiales de Barcelona, ICMAB-CSIC, Campus UAB, Bellaterra 08193, Spain
Email: mcampoy@icmab.es

[d] Institute of Polymer Optoelectronic Materials and Devices, State Key Laboratory of Luminescent Materials and Devices, South China University of Technology, Guangzhou 510640, P. R. China

[e] Department of Materials Science and Engineering, City University of Hong Kong, Tat Chee Avenue, Kowloon, Hong Kong

[f] Sustainability Research Institute, School of Earth and Environment, University of Leeds, Leeds, LS2 9JT

**Keywords:** small-molecule acceptors, organic solar cells, absorption coefficient, machine-learning, density functional theory

## Abstract

Non-fullerene acceptors (NFAs) are excellent light harvesters, yet the origin of such high optical extinction is not well understood. In this work, we investigate the absorption strength of NFAs by building a database of time-dependent density functional theory (TDDFT) calculations of ~500 π-conjugated molecules. The calculations are first validated by comparison with experimental measurements on liquid and solid state using common fullerene and non-fullerene acceptors. We find that the molar extinction coefficient ($\varepsilon_{d,max}$) shows reasonable agreement between calculation in vacuum and experiment for molecules in solution, highlighting the effectiveness of TDDFT for predicting optical properties of organic π-conjugated molecules. We then perform a statistical analysis based on molecular descriptors to identify which features are important in defining the absorption strength. This allows us to identify structural features that are correlated with high absorption strength in NFAs and could be used to guide molecular design: highly absorbing NFAs should possess a planar, linear, and fully conjugated molecular backbone with highly polarisable heteroatoms. We then exploit a random decision forest to draw predictions for $\varepsilon_{d,max}$ using a computational framework based on extended tight-binding Hamiltonians, which shows reasonable predicting accuracy with lower computational cost than TDDFT. This work provides a general understanding of the relationship between molecular structure and absorption strength in π-conjugated organic molecules, including NFAs, while introducing predictive machine-learning models of low computational cost.

## Broader context

The synthetic versatility of organic π-conjugated semiconductors converts them onto the ideal candidates for rational molecular design based on high-throughput screening techniques. Significant advances had been made by trial and error with new but increasingly diverse moieties and materials, primarily using non-fullerene acceptors (NFAs). These have raised the efficiency of organic photovoltaics (OPVs) above 19% in single junctions, to a large extent owing to their high absorption strength. However, the reasons for that remain elusive, thus preventing the molecular tailoring of NFAs with further enhanced light harvesting capabilities that enable breakthrough OPV efficiencies in the years to come. Here we exploit time-dependent density functional theory (TDDFT) calculations performed on NFA molecules and π-conjugated oligomers to investigate what drives their absorption strength higher. The statistical analysis of thousands of molecular descriptors reveals that molecular linearity, planarity, polarizability, and number of π-conjugated carbon atoms correlate strongly with the absorption strength, hence forming a structure-absorption strength relationship that is further exploited to introduce design rules for highly absorbing NFAs. We identify frequent moieties (i.e. molecular fragments) and combinations thereof to drive absorption strength higher in novel NFAs. To speed up the screening of NFA molecular candidates at lower computational cost, we propose exploiting a state-of-the-art machine-learning (ML) model in combination with extended tight-binding Hamiltonians to predict the absorption strength of π-conjugated organic molecules. This work contributes to an improved understanding of the absorption strength of π-conjugated organic molecules while helping the OPV community to design highly absorbing NFAs that maximize the light harvesting capabilities in the green energy market.

**TOC:**

We combine experiments with density functional theory calculations, statistical analysis, and machine-learning to reveal the structure-absorption strength relationship and predict the absorption strength in organic non-fullerene acceptors.

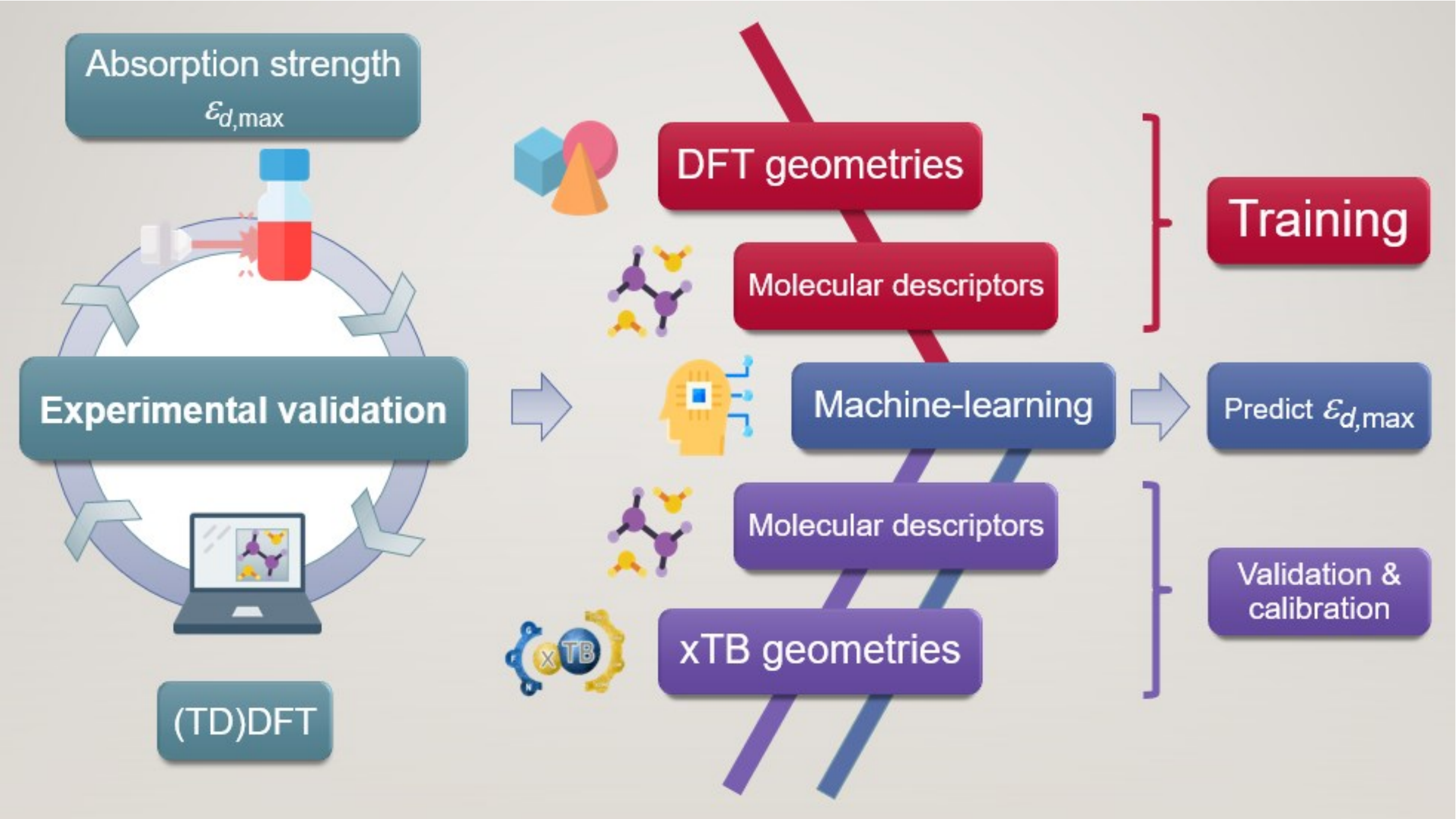

## 1. Introduction

Organic photovoltaic (OPV) energy conversion is a promising option among next generation renewable and sustainable energy technologies for a low-carbon energy future.[1–3] OPV has shown promising potential for various applications, such as indoor photovoltaics (PV),[4–6] semi-transparent solar windows,[7,8] PV greenhouses,[9] and off-grid power supply.[10] Recent OPV devices based on non-fullerene acceptors (NFAs) have demonstrated power conversion efficiencies (PCEs) exceeding 19%,[11,12] much closer to the efficiencies observed in inorganic semiconductor PV technologies such as crystalline silicon and perovskite solar cells, and far higher than values thought attainable in OPV when using fullerene derivatives as the electron-acceptors.[13] The startling progress led by NFAs can be attributed to various advantages over fullerene derivatives, such as band-gap tunability, sharp absorption onset, high emission, high absorption, and low energy losses.[14–16] Among these advantages, the absorption strength of state-of-the-art NFAs is particularly outstanding, as exemplified in **Figure 1**c (a detailed list of chemical names and nomenclatures is provided in **Supplementary Note 1**).[17] For instance, Y6 shows a maximum extinction coefficient ( $\kappa_{max}$ ) over 1.5 in the visible part of the electromagnetic spectrum, as compared to less than 0.75 for fullerene derivatives (PC61BM and PC71BM). High extinction coefficient increases the chance of high quantum efficiency and photogenerated current density, and makes it possible to fabricate highly absorbing OPV films with just a few tens of nanometre-thick photoactive layers. In comparison with workhorse fullerene acceptors, OPV devices based on highly absorbing NFAs could be made comparably thinner than the former, which exponentially raises the output power per weight (i.e. the specific weight in W $g^{-1}$) of OPV devices[18] and might be an effective route toward lower production costs (as less material could employed to achieve an equivalent PCE) and even increase device thermal stability[19]. Moreover, through detailed balance between photon absorption and emission,[20,21] high absorption strength in principle should lead to high emission from the NFAs, while strong NFA emission is believed to be a key reason for NFA-based OPVs to possess low nonradiative voltage losses.[22–26] Despite the clear advantage of strong photo-absorption of NFAs over fullerene derivatives, the phenomenon has attracted much less attention than other properties of NFAs.[22–24,27–30] The features empirically and theoretically proposed[31,32] to lead to strong absorption in π-conjugated polymers are molecular stiffness, linearity, extended π-conjugation and large molecular size. It is therefore of interest to establish whether the same features are associated with strong absorption in NFAs, while seeking design rules for high absorption in future high-performance NFAs.

Excited state calculations based on quantum chemistry methods, such as time-dependent density functional theory (TDDFT),[8,31,33–35] Hartree-Fock method,[36] ab initio Monte Carlo method,[37] second order Møllier-Plesset theory (MP2),[38] and coupled cluster method,[39] have been applied to predict the electronic and optical properties of molecules. Among them, TDDFT is the most widely applied method for excited state calculations, and has shown reasonable accuracy in calculating and predicting the trends in absorption strength of organic molecules,[31,33] as also demonstrated in this work. However, the rapid scaling of computation time with molecular size has been the real obstacle limiting the applicability of TDDFT for excited state calculations on molecules with hundreds of atoms. Given the size and diverse structure of modern NFAs, faster and more efficient methods are therefore needed to establish the relationship between excited-state and molecular properties in NFAs.

The emergence of artificial intelligence (AI) has made it possible to study quantitative structure–property relationships (QSPRs) in molecules with massively improved computational efficiency. As the most popular branch of AI, machine-learning (ML) has attracted much attention in materials science over the last decade, and has been widely applied for material property prediction and material discovery.[40–43] Recently, ML has also gained popularity in OPV scenarios,[44,45,54,46–53] yet existing ML studies related to OPVs have been primarily focused either on the energetics[44,45,55–58] or directly on PCE,[44,49,59–66] with little attention paid to the absorption strength of the photoactive materials.[67,68] Moreover, there are no ML studies explicitly focused on the absorption strength of NFAs beyond the identification of moieties of frequent appearance in highly absorbing molecules.[44] However, QSPR and ML models have been successfully applied to investigate the absorption strength of fluorophores or dyes typically employed in bioimaging, showing encouraging results.[32,69,70] Therefore, it is appealing to apply ML methods in combination with QSPR models to investigate the origin of the large absorption strength in state-of-the-art NFAs.

Here, we present an experimental, TDDFT, QSPR, statistical and ML study of the absorption strength of NFAs to identify the key chemical and structural features that lead to high optical absorption in state-of-the-art NFAs. We exploit a database of nearly 500 unique organic molecules (or 3500 calculations) generated using DFT and TDDFT over several years. We obtain good quantitative agreement between TDDFT calculations of absorption strength and experimental values for state-of-the-art NFAs and fullerenes, which supports the use of TDDFT results for further statistical and QSPR modelling. Accordingly, we extract molecular information from the DFT-optimized geometries by computing nearly 6000 molecular descriptors and first looking for correlations with the absorption strength. The strongest correlations are found between experimentally measured maximum molar extinction coefficient ($\varepsilon_{d,max}$) and two main molecular descriptors from calculations: $\lambda_{1,p}$ and *C2SP2*, which describe the size of the molecule in the direction of maximal atomic polarizability, and the number of $sp^2$ hybridized carbon atoms that are bound to two other carbons (C2), respectively. These quantities can be related to a few key material features leading to high absorption strength: linearity, planarity, and extension of the π-conjugation in the form of fused and closed-ring moieties, in good agreement with previous ML reports on fluorophores and dyes.[32] We further identify several moieties and paired combinations thereof that are frequently found in highly absorbing NFAs, corresponding to thieno[3,2-b]thiophene (TT), thiophene (T), 2-(5,6-difluoro-3-oxo-2,3-dihydro-1H-inden-1-ylidene)malononitrile (2FIC), 2-(3-oxo-2,3-dihydro-1H-inden-1-ylidene)malononitrile (IC) and indaceno[1,2-b:5,6-b′]dithiophene (IDT). These form a catalogue of molecular design rules to further enhance the absorption strength of next-generation NFAs. We then train and test an ensemble learning method, namely a random decision forest (RF), to predict $\varepsilon_{\mathrm{d,max}}$ and provide further information about the most important features in the modelling of absorption strength in NFAs. Finally, we explore the possibility to predict $\varepsilon_{d,max}$ while using a cheaper molecular geometry optimization method based on semiempirical extended tight-binding (xTB) Hamiltonians instead of the expensive DFT approach. We do so by training a RF with our TDDFT database and proving its predictive properties in terms of $\varepsilon_{d,max}$ when interpolated using xTB-optimized geometries. This approach shows application potential in high-throughput screening studies in combination with generative molecular models.

## 2. Results and discussion

### 2.1. Experimental validation of calculated absorption strength using TDDFT

Quantifying how well the TDDFT derived excited state properties agree with the experimental measurements in terms of absorption strength is of utmost importance to validate our theoretical calculations and support further conclusions extracted thereof. Accordingly, we first evaluate the agreement between TDDFT calculations and experimental data in terms of the absorption strength. We compare the absorption strength of a broad catalogue (~10 molecules) of NFA molecules and widely-studied fullerene derivatives (PC61BM and PC71BM, with their molecular structures shown in **Figure 1**a) as obtained from TDDFT calculations, with a variety of optical measurements in both solution and solid state. The measured refractive index ($n$) and extinction coefficient ($\kappa$) of those molecules in thin film obtained using our variable-angle spectroscopic ellipsometry (VASE) measurements are shown in **Figure 1**b and c. Solution state data shown in **Figure 1**d and e are collected from a variety of literature references as detailed in the Supporting database.

As a metric for absorption strength, we initially consider several candidates such as the oscillator strength ($f_{osc}$), the absorption coefficient ($\alpha$) or the imaginary part of the dielectric function ($\varepsilon_2$). In this work, we eventually focus on the maximum molar extinction coefficient ($\varepsilon_{d,max}$, $M^{-1}$ $cm^{-1}$) of NFAs as it shows the best agreement between experimental and theoretical data, as we demonstrate below. $\varepsilon_{d,max}$ constitutes a typical experimental measurement in solution that can also be accessed from myriad literature references. Note that the usual calculations based on single molecules using TDDFT cannot account for solid state effects as they are performed for isolated molecules in vacuum or surrounded by an isotropic medium (such as a solvent using the polarizable-continuum-solvent-model, PCM, **Figure S1**). The derivation of the theoretical $\varepsilon_d$ is provided in the Methods section, which results in a mathematical expression for $\varepsilon_{d,max}$ as

$$\varepsilon_{d,max} = 10\, log_{10}(e)\, N_A \frac{2\pi e\hbar}{3\epsilon_0 m_0 n_r c} f_{osc,max} \frac{1}{\sigma\sqrt{2\pi}}\ , \qquad \text{(Equation 1)}$$

Where $N_A$ is the Avogadro constant, $e$ the elementary charge, $\hbar$ the reduced Planck constant, $\epsilon_0$ the vacuum permittivity, $m_0$ the electron mas, $n_r$ is the refractive index in solution (assumed to be 1.3 of a common organic solvent throughout this study), and c the speed of light. $f_{osc,max}$ is the oscillator strength of the strongest transition among the calculated states within the visible-IR part of the spectrum, and $E_{max}$ is the energy of that transition. The brightest transition is very often the lowest-energy transition in commonly used π-conjugated molecules.[31] We note here that the delta function in Eq. (S13) is replaced with a gaussian distribution function with a peak intensity of $\frac{1}{\sigma\sqrt{2\pi}}$, where $\sigma$ is the gaussian width and assumed to be 0.1 eV for a common organic pi-conjugated molecule.

The experimental $\varepsilon_{d,max}$ from solution can be obtained using the optical density (OD) measurements performed using UV-visible spectroscopy, via

$$\varepsilon_{d,max} = \frac{OD_{max}}{\rho d}, \qquad \text{(Equation 2)}$$

where $OD_{max}$ is the maximum optical density, $\rho$ is the molar concentration (M), and d the light path length of the cuvette (cm). Similarly, the experimental $\varepsilon_{d,max}$ from film can be estimated assuming a mass concentration $\rho_M$ in the film of 1000 g $L^{-1}$ (as a typical value for conjugated polymers and small molecules),[31] either from the maximum absorption coefficient $\alpha_{cm,max}$ ($cm^{-1}$) or extinction coefficient ($\kappa_{max}$) (**Figure 1**c), via

$$\varepsilon_{d,max} = log_{10}(e)\, \alpha_{cm,max} \frac{M_w}{\rho_M} = log_{10}(e) \frac{4\pi\kappa_{max}}{\lambda_{max}} \frac{M_w}{\rho_M}, \quad \text{(Equation 3)}$$

where $M_w$ is the molecular weight in g $mol^{-1}$, and $\lambda_{max}$ the wavelength at $\kappa_{max}$ in centimetre.

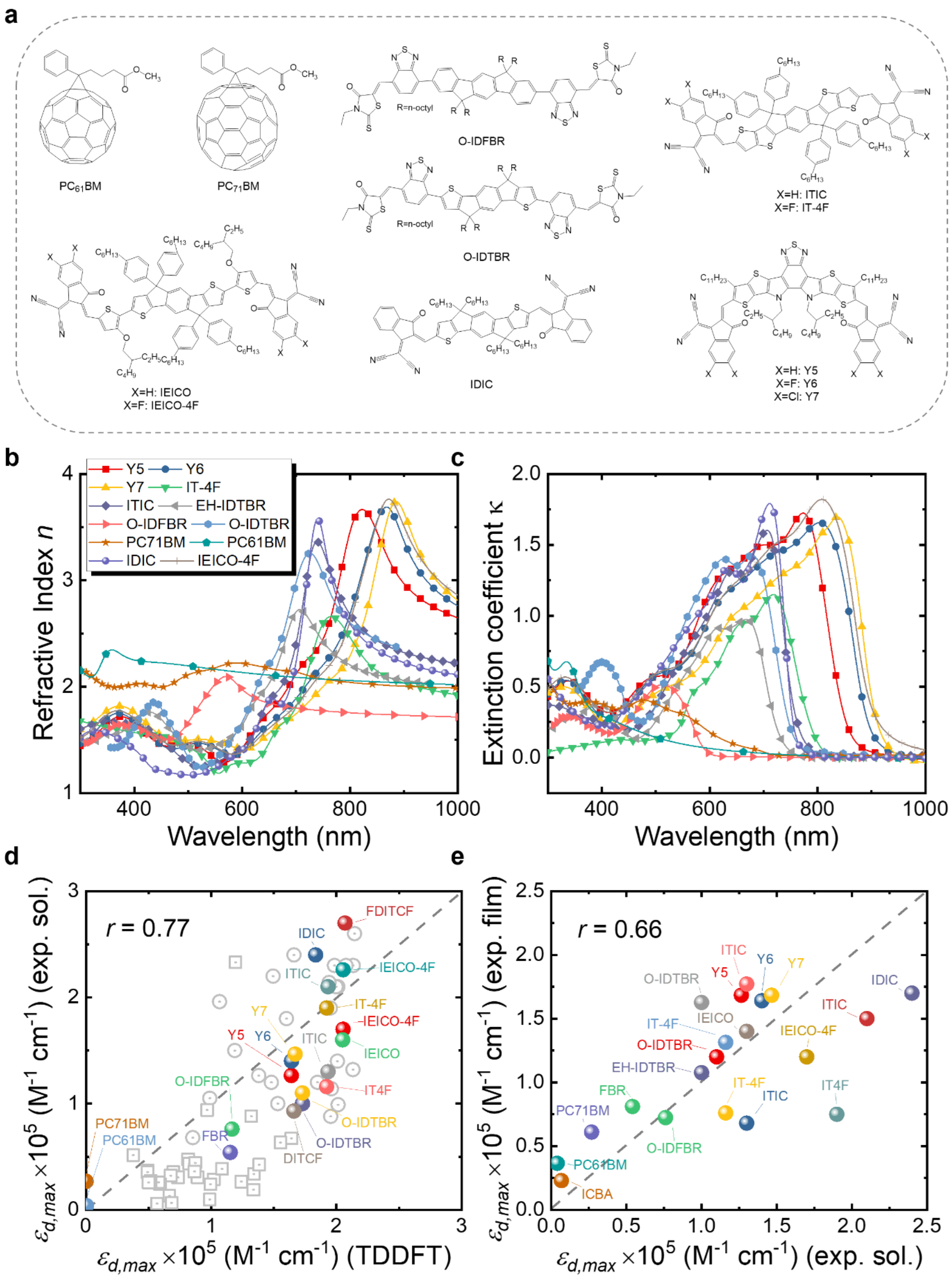

**Figure 1.** a) Molecular structures of typical organic acceptors, including PC61BM, PC71BM, O-IDFBR, O-IDTBR, ITIC, IT-4F, IDIC, IEICO, IEICO-4F, Y5, Y6, and Y7. b) Refractive index and c) extinction coefficient of a larger set of typical organic acceptor thin films measured using VASE. d) Experimental $\varepsilon_{d,max}$ in solution versus calculated $\varepsilon_{d,max}$ in vacuum using

TDDFT of a set of ~80 π-conjugated molecules. e) Estimated experimental $\varepsilon_{d,max}$ in film (solid state) versus that in solution using Eq. (3). Panel (d) contains a subset of well-known NFA molecules that are highlighted in colour. All TDDFT results in panel (d) were performed using the functional B3LYP and basis set 6-311+G(d,p), except for the ones (grey squares) taken from Ref. [71] that are based on the LRC-wPBEh functional and 6-311+G(d) basis set. We also note here that the side chains of molecules are replaced by H atoms or methyl groups in the calculations as they are computationally expensive and do not contribute to the π-conjugation, hence electronic transitions.[31] The experimental data of $\varepsilon_{d,max}$ in film are converted from maximum values of extinction coefficients shown in panel (c) using Eq. (3), while solution data are collected from literature, noting that different values may be present for the same material as retrieved from different sources. Grey dashed lines indicate the perfect match between x and y axis. The data required for generating panels (d) and (e) in this figure are presented in the Supplementary Database.

**Figure 1**d presents the results of the comparison between experimental $\varepsilon_{d,max}$ in solution and theoretical $\varepsilon_{d,max}$ calculated from single molecules using TDDFT in vacuum. A brief discussion of the solvent effect on the absorption strength and the reasons why we choose vacuum medium are provided in **Figure S1**. Despite the scattering of data points, we observe the occurrence of a monotonic relationship between solution and calculated $\varepsilon_{d,max}$ with a Pearson correlation coefficient ($r$) of 0.77. Interestingly, such correlation is no longer observed when quantifying the absorption strength in terms of $\alpha_{max}$ neither when adding further data points from literature on π-conjugated fluorophores to our statistical analysis (**Figure S2**a, r = 0.30), which is believed to be caused by the differences in molecular weight; in that case, only $\varepsilon_{d,max}$ is found to follow a monotonic trend (**Figure S2**b). Some of the material assumptions on refractive index and density required to obtain $\alpha_{max}$ values might be responsible for the observed mismatch. It is worth noting that, expectedly, the correlation between solid state (film) and solution ($r = 0.66$, **Figure 1**e) or calculated $\varepsilon_{d,max}$ ($r = 0.61$, **Figure S3**) is not as good as that from solution data versus calculated $\varepsilon_{d,max}$ ($r = 0.77$, **Figure 1**d, neither for $\alpha_{max}$ as shown in **Figure S4**). Such discrepancy is attributed to solid-state effects such as the aggregation effects,[16] intermolecular orientation,[72,73] and side chain interactions,[74] which are not considered in single molecule excited state calculations.[31] The observed trend that a highly absorbing material in solution will produce highly absorbing films is, nonetheless, generally valid and thus solution data is relevant for devices. Since the NFAs analysed here have a rather similar number of π-electrons ($n_\pi$), the corresponding $\varepsilon_{d,max}$ per π-electron (**Figure S5**) shows

a similar trend as that in **Figure 1**d, **Figure 1**e, and **Figure S3**. Despite the simplicity of single molecule excited state calculations, these data show that using TDDFT calculations of the excited state to deliver $\varepsilon_{d,max}$ can provide a reasonably good approximation to experimental measurements. Moreover, dealing with TDDFT calculations gives us room to correlate key molecular properties, such as molecular size and shape (aspect ratio), linearity, planarity, grafted side chain positions, or functional groups, to the absorption strength using molecular descriptors. These observations provide a foundation from molecular structures to identify the origin and further extend the high optical extinction of NFAs through chemical design rules, as we show in the upcoming sections.

## 2.2. Statistical analysis of the TDDFT absorption strength dataset

The experimental validation of the TDDFT calculations in NFAs supports the use of such results to build an extended database of optimized molecular geometries and excited state properties. The dataset is built by collecting thousands of molecular geometries generated over the last years in our group, making up a total of 3515 calculations on small molecules and oligomers. The distribution of number of atoms in a molecule is shown in **Figure S6** with a majority lying between 50 and 100 atoms. This database is sufficiently diverse to allow us to detect correlations and chemical/structural design rules that could explain and/or further enhance optical absorption in conjugated small molecules.

### 2.2.1. Correlation analysis of molecular descriptors

In the simplest statistical analysis of our TDDFT database, we look for correlations of the absorption strength with respect to a catalogue of molecular descriptors. First, as described in **Supplementary Note 2**, we filter the pristine TDDFT database by identifying duplicate molecules (in terms of molecular weight) and selecting the lowest energy conformer (i.e., optimized geometries in the ground state) among them. As a result, the curated TDDFT database employed in this work consists of 479 π-conjugated small molecules and oligomers with a distribution of moieties shown in **Figure S8**.

Then, we introduce several target features related with absorption strength, starting from the maximum oscillator strength of any calculated transition ($f_{max}$); the maximum oscillator strength of any transition in the visible electromagnetic window (herein constrained between 300-1200 nm or 1-4 eV for its relevance in solar energy harvesting applications) ($f_{max,vis}$); and the sum of oscillator strengths of all transitions in the visible window, $f_{sum,vis}$. These three features are also evaluated per $n_\pi$ for the molecule, i.e., $f_{max}/n_\pi$, $f_{max,vis}/n_\pi$ and $f_{sum,vis}/n_\pi$. We then consider the maximum absorption coefficient ($\alpha_{max}$) obtained using Eq. (1) and Eq. (3); the maximum of the imaginary part of the dielectric function ($\varepsilon_{2,max}$);[31] and $\varepsilon_{d,max}$. Finally, we compute the spectral overlap between the OD ($d\alpha(E)$, where $d$ is set to a typical film thickness value of 100 nm and $\alpha(E)$ derives from the Gaussian-broadened spectrum of $f$ in the visible spectral range taking a standard deviation of 0.1 eV) and the AM1.5G solar photon flux spectrum ($\Phi_{\mathrm{AM1.5G}}$), namely $f_{overlap} = \frac{\int_{1\,eV}^{4\,eV} \Phi_{AM1.5G}(E) d\alpha(E) dE}{\int_{1\,eV}^{4\,eV} \Phi_{AM1.5G}(E) dE}$.

These features, together with their corresponding histograms (**Figure S13**) in terms of Spearman's rank correlation coefficients ($\rho$), are explained in more detail in **Supplementary Note 2**. Molecular descriptors are calculated using up to four different open-source packages[75–78] (**Supplementary Note 2**) to generate a (curated) collection of 3239 entries (including 40 electronic descriptors derived from the TDDFT calculations, namely the energy of the molecular orbitals ranging from HOMO-19 to LUMO+19). Then, we scan for statistical correlations between those descriptors and all target features introduced above, from which we consider as highly correlated descriptors those showing $\rho \geq 0.7$ as threshold. However, since some descriptors are calculated in groups or families where weighting factors are varied among atomic masses, van der Waals volumes, electronegativities, ionization potentials or polarizabilities, we usually encounter sets of multicollinear descriptors that show very similar trends with respect to the target feature. Accordingly, to drop redundant (collinear) descriptors we classify them into clusters to select the most representative candidate of each bundle (i.e., cluster). This serves us to simplify the identification of characteristic and well-correlated descriptors families. The clustering algorithm applied to analyse multicollinear descriptors based on ρ and r values is further described in **Supplementary Note 3**.

After running the clusterization of descriptors on all target features, we identify strong correlations with molecular descriptors for $f_{max}$, $f_{sum,vis}$ and $\varepsilon_{d,max}$ (i.e., implicitly $f_{max,vis}$). For the remaining target variables ($f_{overlap}$, $\alpha_{max}$, $\varepsilon_{2,max}$, $f_{max}/n_\pi$, $f_{max,vis}/n_\pi$ and $f_{sum,vis}/n_\pi$), we do not identify molecular descriptors with ρ above the threshold value (0.7) and they are generally below 0.6 units, see **Figure S13**. The lack of correlation for $f_{overlap}$ could be justified by the existence of a gas-to-solid shift in the corresponding absorption spectrum, which prevents proper matching of the Gaussian-broadened absorption features with the solar photon flux. Regarding $\alpha_{max}$ and $\varepsilon_{2,max}$, the estimation of these values from TDDFT calculations requires taking generalized assumptions on several materials properties (such as density or refractive index) that might be enough to disturb the underlying trends in our heterogeneous material database. For the quantities normalised by the number of pi electrons, i.e. $f_{max}/n_{\pi}$, $f_{max,vis}/n_{\pi}$ and $f_{sum,vis}/n_{\pi}$, the weak correlation is expected since normalization tends to deviate from linear correlations depending on the straightness of the molecule.[31] Due to the strong correlation between size of the molecule and oscillator strength as discussed below based on C2SP2, the normalised quantity is believed to be a secondary factor, therefore not clear correlations are observed. In the successful correlation cases (i.e. $f_{max}$, $f_{sum,vis}$ and $\varepsilon_{d,max}$) and with the given thresholds of 0.7 units for $\rho$ and $r$, we identify a single feature cluster lead by the $\lambda_{1,p}$ descriptor in the case of $f_{max}$ and $\varepsilon_{d,max}$ (**Figure 2**a). For $f_{sum,vis}$, a threshold $\rho$ of 0.68 reveals C2SP2 as a rather descriptive molecular feature (**Figure 2**b). Interestingly, C2SP2 is also found in the main cluster represented by $\lambda_{1,p}$ in $f_{max}$ and $\varepsilon_{d,max}$, and we could not identify any strong correlations between the absorption strength (in any of its proposed metrics) and electronic descriptors (from HOMO-19 to LUMO+19 energy levels). Note that $\varepsilon_{d,max}$ values in excess of $2.5 \times 10^5$ $M^{-1}$ $cm^{-1}$ in **Figure 2**a and b are mostly attributed to artificially straight conjugated oligomers with >10 monomers contained in our database, for which the straightness, hence high $\varepsilon_{d,max}$, are unlikely to be maintained in the experimental solid state scenario. In fact, only the exemplary and asymmetric NFA known as BDTP-4F (inset of **Figure 2**a)[79,80] surpasses that threshold with a record $\varepsilon_{d,max}$ in our NFA dataset ($2.7 \times 10^5$ $M^{-1}$ $cm^{-1}$, and $2.4 \times 10^5$ $M^{-1}$ $cm^{-1}$ measured in $CHCl_3$ solution).[79]

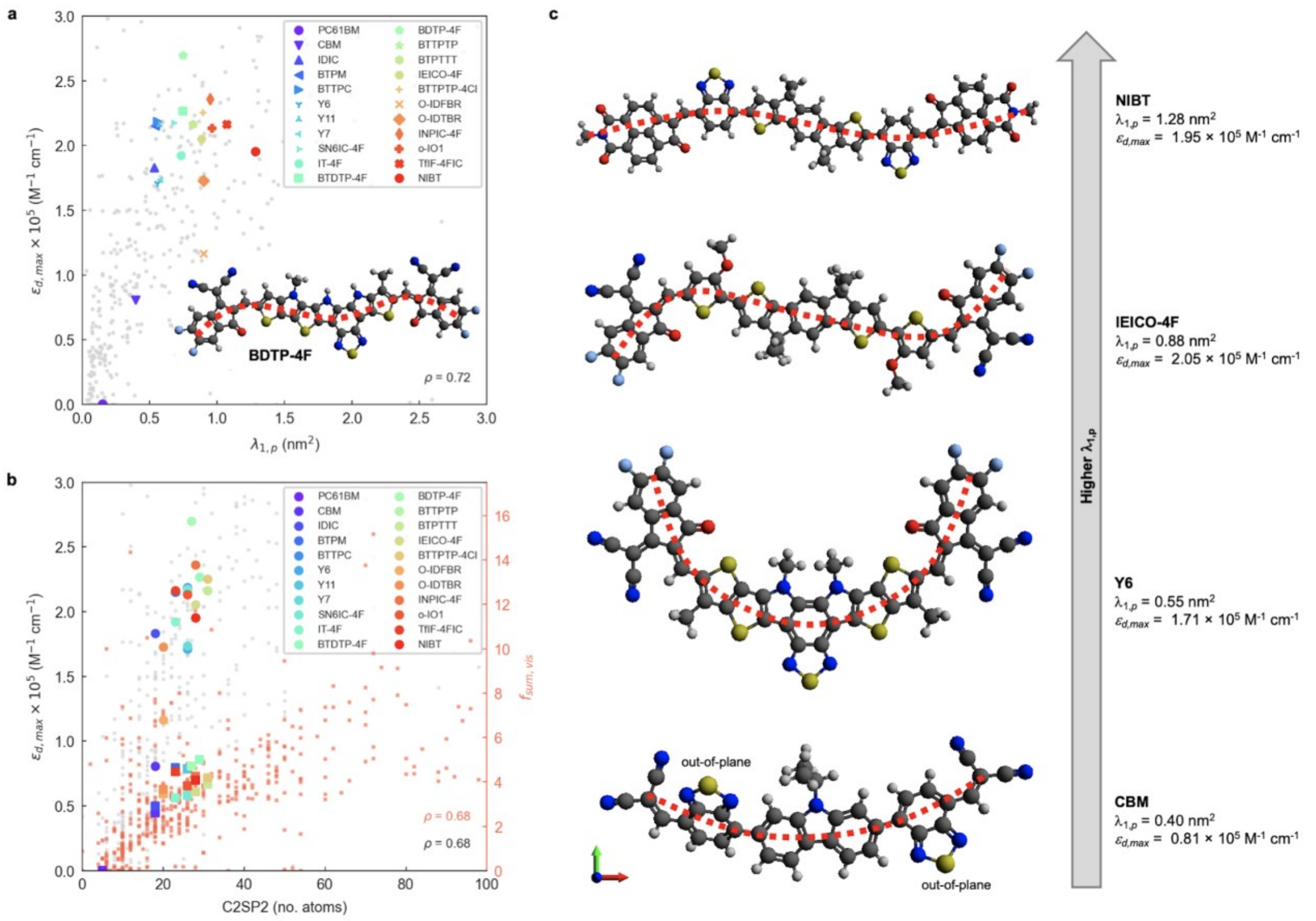

**Figure 2.** (a) Correlation between $\varepsilon_{d,max}$, as calculated from TDDFT, and $\lambda_{1,p}$ as obtained in the database of 479 molecules. The DFT-optimized geometry of BDTP-4F is shown in the inset. (b) Correlation between $\varepsilon_{d,max}$ (and $f_{sum,vis}$ in the secondary axis) and C2SP2 in that same database. (c) DFT-optimized geometries of archetypal NFAs ordered by increased values of $\lambda_{1,p}$ from bottom to top (CBM < Y6 < IEICO-4F < NIBT). Dotted red lines tentatively indicate the overall curvature of the main conjugated backbone of the molecule. $\lambda_{1,p}$ and C2SP2 describe the size of the molecule in the direction of maximal atomic polarizability, and the number of doubly bound carbon atoms ($sp^2$ hybridized) bound to two other carbons (C2), respectively.

$\lambda_{1,p}$ is part of a bundle of three-dimensional molecular size and shape descriptors known as weighted holistic invariant molecular (WHIM) descriptors.[81–83] These can be interpreted as a generalized search for the principal axes with respect to a defined atomic property.[84] In this particular case, $\lambda_{1,p}$ is obtained by performing a principal component analysis (PCA) on the centred atomic coordinates of the molecule using a covariance matrix ($s_{jk}$) that is weighted by the atomic polarizabilities ($p_i$):

$$s_{jk} = \frac{\sum_{i=1}^{A} p_i (q_{ij} - \overline{q_j})(q_{ik} - \overline{q_k})}{\sum_{i=1}^{A} p_i}, \qquad \text{(Equation 4)}$$

where $s_{jk}$ is the weighted covariance between the jth and kth atomic coordinates; $A$ is the total number of atoms; $p_i$ is the (tabulated) polarizability of the ith atom; $q_{ij}$ and $q_{ik}$ represent the jth and kth coordinate of the ith atom $(j, k = x, y, z)$, respectively; and $\bar{q}$ is their average value.[84] After diagonalization of the polarizability-weighted covariance matrix, the first eigenvalue ($\lambda_{1,p}$) quantifies the size of the molecule in the direction of maximal polarizability variance. Interestingly, the third eigenvalue ($\lambda_{3,p}$) approaches zero in planar molecules as a result of absence of variance in the out-of-plane (z) direction.[84] On the other hand, C2SP2, which is not in the WHIM group, accounts for the number of doubly bound carbon atoms ($sp^2$ hybridized, SP2) bound to two other carbons (C2), thus constituting a two-dimensional descriptor of fast computation. The correlation between C2SP2 and absorption strength can be relatively easier to understand, as C2SP2 to some extent represents the size of the conjugated molecule. Enlarging the size of the molecule increases the total number of π-electrons, which controls the total oscillator strength following the Thomas-Reiche-Kuhn rule. For the molecules that are extended along one direction, such as linear oligomers, increasing the size should enhance the oscillator strength of the first transition,[31] i.e. the dominant one.

To further interpret these two magnitudes ($\lambda_{1,p}$ and C2SP2) as the main correlated descriptors with $\varepsilon_{d,max}$ and $f_{sum,vis}$, we inspect the DFT-optimized geometries of archetypal NFAs (**Figure 2**c). The observed trend suggests that optical extinction monotonically increases with $\lambda_{1,p}$ (**Figure 2**a) in molecules having most of their polarizable atoms arranged along a main axis, i.e., linear molecules. While CBM shows large torsion angles mainly affecting the 2,1,3-benzothiadiazole (BT) moieties (thus making the molecule non-planar and increasing $\lambda_{3,p}$, see **Figure S14**), Y6 shows a characteristic curved geometry that limits its $\varepsilon_{d,max}$ despite showing improved planarity. The NFA with the highest $\lambda_{1,p}$ (NIBT) shows both linearity and planarity, with most of the more polarizable atoms (mainly C and S) lying along the principal polarizable axis of the molecule. Thus, in terms of molecular geometry, the absorption strength of NFAs could be further enhanced by distributing most of the atomic polarizability along a main axis while keeping good planarity and minimizing curvature. However, $\lambda_{1,p}$ is not the sole molecular descriptor governing absorption strength, as BDTP-4F shows ca. 40% lower $\lambda_{1,p}$ (0.75 $nm^2$) yet ca. 40% higher $\varepsilon_{d,max}$ than NIBT (**Figure 2**a), which suggests that the molecular symmetry of NFAs could be another important factor affecting $\varepsilon_{d,max}$. Our preliminary investigations on this issue indicate that molecular asymmetry, as quantified by the WHIM symmetry index $G_u$, might drive absorption strength higher (**Figure S15**a), yet we require a larger NFA database including more asymmetric molecules to further explore such an observation. Also, we acknowledge that this observation might be biased by the systematic omission of side chains in the TDDFT calculations. By comparing $\lambda_{1,p}$ in a selection of small molecule acceptors geometrically optimized with and without side chains (**Figure S16**a), we observe that in most cases the addition of side chains either decreases $\lambda_{1,p}$ slightly or keeps it invariant. Still, the positive correlation of $\lambda_{1,p}$ with respect to $\varepsilon_{d,max}$ is maintained (**Figure S16**b). Furthermore, the presence of napthalene imide derivatives in the molecular structure of NIBT could be hindering further increase of the absorption strength with $\lambda_{1,p}$, as suggested by our statistical analysis of frequent moieties in the selection of good light harvesters (presented in the next section). On the other hand, an increase of $n_\pi$ in the molecule in the form of closed-ring conjugated moieties will systematically increase C2SP2 and accordingly $f_{sum,vis}$. These findings support the previously known design rules in terms of molecular linearity and π-conjugation enabling large oscillator strength in organic small molecules and polymers, and are consistent with a recent study on chromophores.[32] In particular, trans- conjugated polymer stereoisomers are known to possess higher optical extinction due to their increased straightness and persistence length,[31] which agrees with our observations on exemplary curved (Y6) and more linear (NIBT) NFAs.

The energy of the first optical transition ($E_1$) is also of practical importance in light harvesters such as NFAs as the lower energy part of the solar spectrum, down to ~1 eV, contains a higher photon flux density. Our results show the number of heteroatoms in the molecule as the most correlated feature with $E_1$ ($\rho$ = -0.72, **Figure S17**a) while forming a single feature cluster, yet neither $\lambda_{1,p}$ nor C2SP2 show strong correlations with $E_1$. This fact prevents the introduction of molecular design rules targeted at $E_1$ using $\lambda_{1,p}$ or C2SP2. However, we acknowledge a negative correlation between $E_1$ and $f_{osc,max}$ among common NFAs that suggests further room for absorption strength increase as $E_1$ is reduced (**Figure S18**).

### 2.2.2. Chemical insights into highly absorbing molecules

Beyond molecular descriptors, we investigate the relationship between the choice of moieties and absorption strength to provide further material design rules for highly absorbing conjugated small molecules. Our objective is to identify overrepresented moieties in the subset of high-absorbing molecules (which we arbitrarily define as those having $f_{osc,max}$ > 2.5, thus setting a population of size p) with respect to the entire molecular dataset (population of size P). Accordingly, we identify the molecular motifs present in the molecules by comparing their structures (as derived from SMILES notation) with those of a previously built database of moieties (also SMILES-based). This database of moieties was partly inherited from a previous work[44] and extended with further motifs present in our particular dataset (see **Supplementary Note 4** and the spreadsheet included as Supplementary database). Afterwards, we consider that a discrete hypergeometric distribution is adequate to model our molecular dataset and the fragments found therein[44] to calculate the corresponding Z-scores as $Z = (k - \bar{k})/\sigma_k$, where $k$ is the number of high-absorbing molecules containing certain moiety; $\bar{k}$ is its expected value, defined as $pK/P$ where $K$ corresponds to the number of molecules in the entire dataset containing that same moiety; and $\sigma_k = \sqrt{pK(P-K)(P-p)/(P^2(P-1))}$ is the standard deviation of the hypergeometric distribution. Z-scores will indicate (in units of $\sigma_k$) which moieties are overrepresented or underrepresented in the subset of high-performing molecules with respect to the expected values when looking at the entire dataset. Our results (**Figure 3**) suggest that thieno[3,2-b]thiophene (TT), thiophene (T), 2-(5,6-difluoro-3-oxo-2,3-dihydro-1H-inden-1-ylidene)malononitrile (2FIC), 2-(3-oxo-2,3-dihydro-1H-inden-1-ylidene)malononitrile (IC), indaceno[1,2-b:5,6-b′]dithiophene (IDT), 2-methylene malononitrile, cyanide, and aniline are particularly frequent in highly absorbing molecules. Interestingly, four of those molecular fragments (TT, T, 2FIC and IC) are contained in the chemical structure of the workhorse NFA Y6 (**Figure 2**c). Contrarily, napthalene imide derivatives, as typically encountered in n-type small molecules and conjugated polymers; 4H,8H-benzo[1,2-c:4,5-c']dithiophene-4,8-dione and benzo[1,2-b:4,5-b']dithiophene fragments are mostly underrepresented in the selection of high-performing light harvesters.

We further study the existing correlation between pairs of moieties to understand in which way the different molecular fragments should (or should not) be combined to retrieve highly-absorbing molecules. Our analysis starts by creating molecular subsets determined by the presence of a given moiety, which acts as source node (coloured in black) in the network graph shown in **Figure 3b.** Within that subset, we identify the high-absorbing molecules ($f_{\mathrm{osc,max}} > 2.5$) and compute the Z-scores of their moieties (child nodes, coloured in grey in **Figure 3b**) with respect to the molecules of the entire molecular subset. As per the network shown in **Figure 3b**, the absolute Z-scores will determine the width of the edges connecting the nodes (moieties) and its sign the colour of the edge (green for positive Z-score [overrepresentation] and red for negative Z-score [underrepresentation]). Therefore, green and thick edges connect pairs of molecules that are more frequently found in high-absorbing molecules whereas thick and red edges indicate combinations of moieties that lead to less absorbing molecules. In this analysis, we set up 8 different source nodes corresponding to the most overrepresented moieties observed in **Figure 3a**. As a result, **Figure 3b** can be interpreted as a catalogue of design rules relating pairs of moieties with high oscillator strength in π-conjugated small molecules.

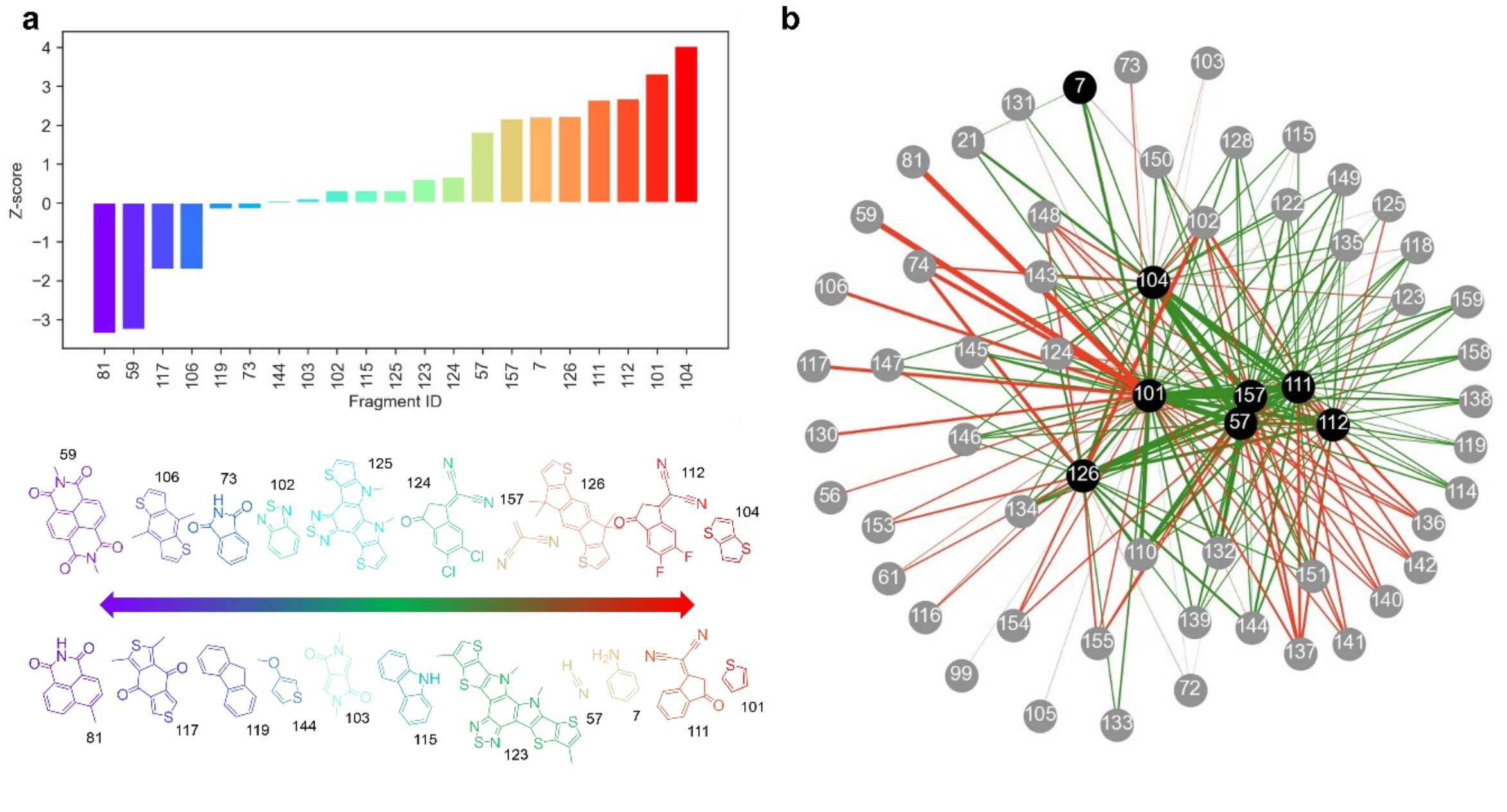

**Figure 3**. (a) Z-scores obtained from the discrete hypergeometric distribution of moieties in the highly-absorbing molecules ($f_{osc,max} > 2.5$) with respect to the entire molecular dataset, for moieties activated at least 10 times. The corresponding structures of identified moieties are shown. (b) Network graph of Z-scores relating pairs of moieties. Source nodes are coloured in black whereas child nodes are coloured in grey. The colour of the edges corresponds to the sign of the Z-score (green for positive, red for negative). The width of the edges scales with the absolute value of the Z-score.

### 2.3. Machine-learning modelling of the absorption strength

Besides providing useful chemical insights from a material design perspective, molecular descriptors can be exploited to feed regression models and draw predictions on certain target features, forming the so-called quantitative structure-property relationship (QSPR) and quantitative structure-activity relationship (QSAR) models.[81,82,84,85] In the present study, we train and test several ML models fed with molecular and electronic descriptors obtained from TDDFT calculations to predict the value of $\varepsilon_{d,max}$ in conjugated small molecules and oligomers. Finally, we propose exploiting such ML model (trained with TDDFT data) to predict $\varepsilon_{d,max}$ in molecules optimized using a semi-empirical quantum chemistry method, i.e. xTB.[86] This renders possible thanks to the geometrical similarity of the TDDFT and xTB ground state conformers, which lead to similar (geometrical) descriptors values; and the calibration of their corresponding energy levels, as per the required inputs of the ML model herein employed. Therefore, further molecular candidates beyond the pristine dataset could be geometrically optimized using solely xTB Hamiltonians and their absorption strength predicted using such ML model. This approach effectively bypasses the use of TDDFT calculations when screening the absorption strength of novel molecules, which results in less demanding computations and higher throughput. The present ML workflow will open the possibility to accelerate the screening of high-performing molecular candidates with low-to-moderate computational requirements (further discussed in **Supplementary Note 5**).

2.3.1. Modelling $\varepsilon_{d,max}$ with random decision forests

From the analysis of descriptors shown in Section 2.2.1, we identified two main feature clusters represented by $\lambda_{1,p}$ and C2SP2. We tentatively consider these two descriptors as independent variables in baseline models (such as 1-nearest neighbour and linear regression) targeted to $\varepsilon_{d,max}$. For the model training and testing, we split our pristine dataset onto two subsets, namely the training set (gathering 70% of the data, randomly selected) and the testing set (gathering the remaining 30% of the data). Such baseline models are picked according to a recently introduced catalogue of good practices in the ML field,[87] to demonstrate the requirement of more advanced regressors (namely ML) in successful data modelling. The models are scored and quantitatively compared based on workhorse fitting metrics, such as their coefficient of determination ($R^2$); their adjusted coefficient of determination ($R^2_{adj}$, which adds penalties as the number of parameters increases, see **Supplementary Note 2**); and their Pearson correlation coefficient (r), as retrieved in the training (fitting) and test sets. The inherent mathematical simplicity of the baseline models results in poor fitting scorings (**Figure S19** and **Table S1**) yet they suggest that feature selection procedures could end up in higher-performing models.

Accordingly, we deploy a state-of-the-art ML method, namely a RF, to aid in both aspects: feature selection and building of $\varepsilon_{d,max}$ models of higher accuracy. RFs constitute one of the simplest and most widely applied ML methods in molecular screening and data mining studies.[32,45,47,88] They are particularly appealing for their straightforward implementation through open-source Python libraries such as Scikit-Learn,[89] and also for their inherent robustness against overfitting and fast optimization. RFs are formed by an ensemble of decision trees (estimators) that are executed in parallel and independently from each other. Decision trees serve to classify data by starting from a single root node that is subsequently divided into child nodes, the latter being chosen randomly among the input features. At every node splitting step (i.e., decision making), the algorithm selects the pathway that minimizes the mean square error (MSE). Eventually, when every tree reaches its maximum extension (which is set arbitrarily via model hyperparameters), the predictions of all trees are averaged (ensembled), hence constituting the final predicted value of the RF. At this stage, myriad cross-validation (CV) techniques exist to evaluate the quality of the model and help in the tuning of hyperparameters. CV methods can estimate the ML model performance, evaluate potential over- or underfitting, and quantify how accurate the model is on drawing predictions on unseen data. In this work, we adopt two common cross-validation schemes, namely a repeated holdout CV; and a leave-one-out cross-validation (LOOCV). On the one hand, in a repeated holdout CV the pristine dataset is randomly split onto two distinct subsets, namely the training (here gathering 70% of the data) and testing (the remaining fraction of data, i.e. 30%) subsets. The model is trained and tested on the respective subsets, and the corresponding statistical metrics ($R^2$, r, MSE, etc.) annotated. Eventually, the process is repeated k times (10-fold in this work), and all metrics are averaged to evaluate the ML model performance (its CV score). On the other hand, in a LOOCV the holdout process is taken to the extreme as the testing subset consists of a single data point while the remaining data is used in the training step. The process runs recursively for all data, thus eventually all data points are used for training and testing in the LOOCV protocol. Yet being computationally expensive, a LOOCV results in a more accurate estimate of model performance.

**Table S1** includes the performance of an out-of-the-box RF model trained and cross-validated using 300 trees (estimators). Exemplary comparisons between the two previous baseline models (1-nearest neighbor and linear regression) and the out-of-the-box RF model are found in **Figure S19**. The RF models indicate that scoring functions ($R^2, r$) well above 0.6-0.8 are feasible upon careful feature selection and further optimization of the RF regressor. Feature selection in RFs is usually performed by filtering variables based on their feature importance, which is a metric that accounts for how much a feature decreases the weighted variance in the node splitting steps of the decision trees. This property enables feature ranking to then apply myriad algorithms to filter out the least important variables as seen by the RF regressor. In this work, we perform a recursive feature elimination (RFE) procedure to the initial library of 3239 descriptors as described in **Supplementary Note 2**. In a RFE protocol, a significant fraction of the initial population of features is dropped in successive training steps of the RF ensemble. Features are dropped based on their corresponding feature importance until reaching an arbitrarily low number of input variables, hence simplifying the original model. Our RFE analysis shows that a threshold average $R^2$ of 0.70 is achieved using a 12-variable model ($R^2 = 0.70 \pm 0.05$, $r =$

*0.84 ± 0.03*), which outperforms the RF model presented earlier while including a drastic reduction in the number of variables (from 3239 to 12). The sweet spot in model accuracy and number of degrees of freedom is found for the 10-variable model, which shows the maximum average $R_{adj}^2$ (0.67 ± 0.06).

Notably, a threshold $R^2$ of 0.60 is already achieved training a 3-parameter RF model ($R^2 = 0.63 \pm 0.06, R_{adj}^2 = 0.62 \pm 0.06, r = 0.80 \pm 0.03$), which is particularly appealing given its simplicity. The resulting three-variable model includes one three-dimensional descriptor ($\lambda_{1,v}$ or WHIM_45, as computed by the RDKit library, **Figure 4**a), one two-dimensional descriptor (CIC3, as computed by PaDEL software, **Figure 4**b) and one electronic descriptor, in this case the energy level of the second molecular orbital below the frontier HOMO (HOMO-2, **Figure 4**c). $\lambda_{1,v}$ refers to the first eigenvalue of the covariance matrix weighted by the atomic van der Waals volumes; thus, $\lambda_{1,v}$ is included in the multicollinear feature cluster represented by $\lambda_{1,p}$ that we previously and statistically identified, showing nearly perfect correlation ($r = 0.99$) with $\lambda_{1,p}$. Accordingly, $\lambda_{1,v}$ can be exchanged by $\lambda_{1,p}$ without loss of performance in the RF model. This finding confirms that the linearity of the molecule (either quantified in terms of polarizabilities or van der Waals volumes) plays a key role in determining its absorption strength in the form of $\varepsilon_{d,max}$. On the other hand, CIC3 is a graph-based, third-order neighbourhood symmetry index[84] which lacks a straightforward interpretation due to its mathematical complexity. We observe, however, that it linearly scales as $\log_2$ A, with A being the total number of vertices (atoms) in the graph (molecule)[84] thus likely reflecting the size of π-conjugation as per the characteristics of our dataset. The interpretation of HOMO-2 as an important descriptor is more challenging, and it is not possible to substitute it by a different descriptor without a noticeable drop in the model performance (excepting HOMO-1, which shows r = 0.96).

Interestingly, electronic descriptors (in particular) are required for the RF models to achieve their highest potential and scoring despite we have not observed strong correlations in our earlier statistical analysis. To probe it, we have performed the same RFE protocol yet skipping the set of electronic descriptors among the input features. Our results show that the top performing RF models (selecting 29 variables and getting $R^2 = 0.58 \pm 0.06$, $R_{adj}^2 = 0.48 \pm 0.07, r = 0.78 \pm 0.04$; or selecting 9 variables to obtain $R_{adj}^2 = 0.52 \pm 0.06$, see **Figure S12**) are yet behind the scorings recorded when the electronic descriptors are included in the list of features. Note that the performance without electronic descriptors is lower than the 3-parameter model that includes HOMO-2 as descriptor, highlighting its positive effect on the performance of the RF regressor.

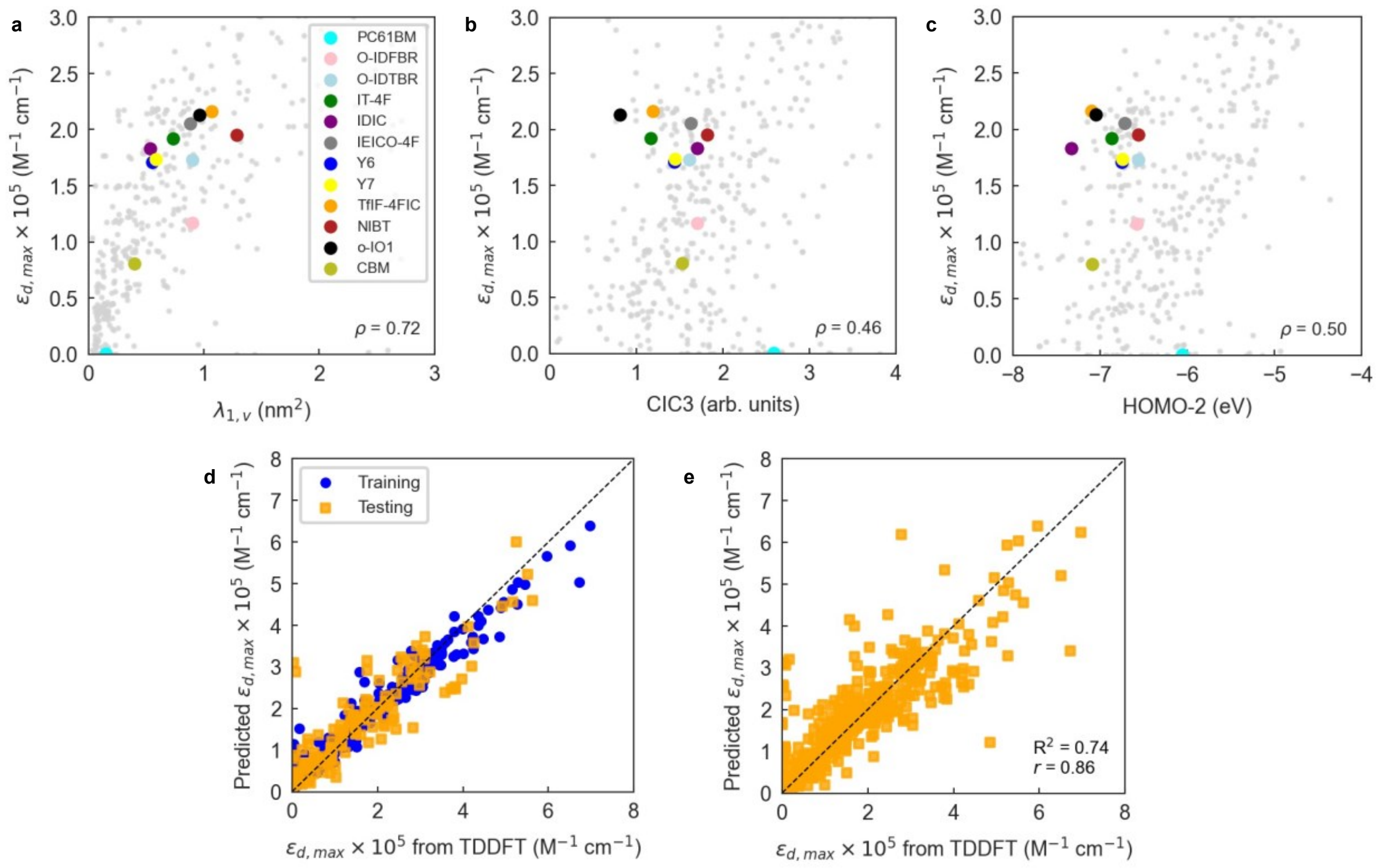

**Figure 4.** Correlation plots for $\varepsilon_{d,max}$ and the three most important descriptors retrieved by the RF model: (a) $\lambda_{1,v}$; (b) CIC3; and (c) HOMO-2. (d) Holdout cross-validation run of a RF ensemble to predict $\varepsilon_{d,max}$. 70% of the data is randomly selected for training and the remaining fraction is used for testing; the process is repeated 10 times and the statistical metrics averaged. The RF model is trained with three molecular descriptors ($\lambda_{1,v}$; CIC3; and HOMO-2) and a Morgan fingerprint vector of 64 bits. (e) Leave-one-out cross-validation (LOOCV) of that same RF model using the optimized hyperparameter of 1200 estimators.

Molecular fingerprints have also been extensively exploited as input vectorial descriptors in statistical and ML models focused on feature prediction.[44,90–92] Molecular fingerprints are usually represented as bit activation vectors of arbitrary length and degree of complexity, representing the absence or presence of certain molecular (bonding) pattern, moiety, functional group, or atom. In this work, we exploit the RDKit library to generate moiety fingerprints, MACCS keys, Morgan fingerprints, path-based or topological fingerprints, E-state fingerprints, and Coulomb vectors. These fingerprints are quickly computed and serve to complement and improve the learning process of the ML models employed herein.

To better analyse the influence of the different fingerprint vectors in improving the RF scoring, we trained and cross-validated the 3-parameter RF model previously found in combination with all fingerprint vectors generated. The results shown in **Table S2** indicate that by adding a Morgan fingerprint vector of 64 bits to the initial set of input features the model performance can be substantially improved: $R^2$ increases by 10% (relative), and r by another (relative) 5% (see **Figure 4**d). Therefore, Morgan fingerprints are particularly suitable to fine-tune the training and prediction accuracy of $\varepsilon_{d,max}$ in RF models although lacking of a straightforward physical interpretation. Additional refinement of the RF hyperparameters results in further improved models. We performed this optimization through a randomized search (in 350 iterations) of the hyperparameters controlling the number of estimators in the RF, the minimum number of samples per leaf node and the minimum number of samples required to split an internal node, which constitute the main adjustable hyperparameters of the RF algorithm. These results are shown in **Table S3**, together with the scoring obtained in a rigorous LOOCV of the optimized RF model (**Figure 4**e). As an alternative ensemble of decision trees, we have also tested and optimized an Extra Trees (ET) regressor in Scikit-Learn. Its performance is, however, very close to that attained in the workhorse RF regressor (**Table S3** and **Figure S20**).

2.3.2. Bypassing TDDFT calculations through machine-learning and extended tight-binding xTB Hamiltonians have recently emerged as semi-empirical and low computational cost quantum chemistry methods.[86] These have a remarkable potential in molecular screening when implemented in multilevel workflows where xTB is exploited first to identify plausible candidates using a minimal fraction of computational resources, to then leave room for higher-level DFT methods in selected candidates.[86] In this work, we propose exploiting a ML model trained with DFT data to predict $\varepsilon_{d,max}$ in molecular geometries optimized using xTB (**Figure 5**a). This is expected to enable faster molecular screening and geometrical optimization steps, as both being entirely run using xTB Hamiltonians; followed by absorption strength ($\varepsilon_{d,max}$) prediction in a TDDFT-trained RF model. Notably, our estimations show that the geometrical optimization step using GFN2-xTB is ca. 3000 times faster than using DFT with a hybrid functional (B3LYP/6-311+G(d,p)), as discussed in **Supplementary Note 5** and **Table S4**.

Nevertheless, the dissimilarity between xTB- and DFT-optimized molecular geometries might have a direct impact on the value of the (three-dimensional) molecular descriptors, and hence on the final accuracy of the interpolated ML model if some of those are included. Accordingly, we have first quantitatively compared both sets of molecular (non-electronic) descriptors by computing r in all of them and found that the median of their distributions is very close to unity in all cases (**Figure S21**). Based on this finding, we proceed by training the RF model with TDDFT-derived descriptors and exploring how well the model interpolates when fed with xTB-derived descriptors. **Figure S22**a shows a leave-one-out interpolation of a RF model trained using TDDFT data and interpolated on GFN2-xTB-optimized molecules, descriptors and energy levels.[86,93,94] In this kind of model validation, all TDDFT data is used in the training step excepting that for a single molecule, for which we retrieve its corresponding xTB-optimized geometry and descriptors as the sole interpolation (testing) dataset; this procedure is subsequently repeated for all molecules. Thus, the model performance is assessed by comparing the actual TDDFT-derived $\varepsilon_{d,max}$ of the molecules (x-axis in **Figure 5**b) with that predicted by a RF model trained with TDDFT data and interpolated using xTB-derived descriptors (y-axis in **Figure 5**b). This is useful to evaluate whether such RF model fed with TDDFT data could be exploited to predict $\varepsilon_{d,max}$ in unseen molecules that are geometrically optimized through xTB Hamiltonians.

Our first model takes as inputs the three molecular descriptors found previously to be the most important features in the RF model together with their corresponding (64-bit) Morgan fingerprints. The scoring of the LOOCV in this preliminary model ($R^2 = 0.53, r = 0.74$) is limited due to the existence of a mismatch between the absolute energy levels retrieved by either DFT (B3LYP) or GFN2-xTB methods (**Figure S22**b). Thus, the RF model trained on TDDFT data needs proper calibration of the energy levels obtained through GFN2-xTB, which we perform using either a linear regression, a support vector regressor (SVR) or an additional RF model (**Figure S22**c). By applying such calibration on the HOMO-2 energy levels, we obtain the champion RF model ($R^2 = 0.61, r = 0.78$) shown in **Figure 5**b using three molecular descriptors and a 64-bit Morgan fingerprint vector. Hence, **Figure 5**b shows that molecular databases of xTB-optimized geometries could be exploited in combination with TDDFT-trained ML models to predict the absorption strength ( $\varepsilon_{d,max}$ ) at significantly lower computational cost and with reasonable accuracy. The statistical analysis and ML modelling framework introduced here is thus expected to show large potential in the high-throughput screening of highly absorbing molecular candidates in combination with generative models (autoencoders and neural networks) as part of future work in the group.

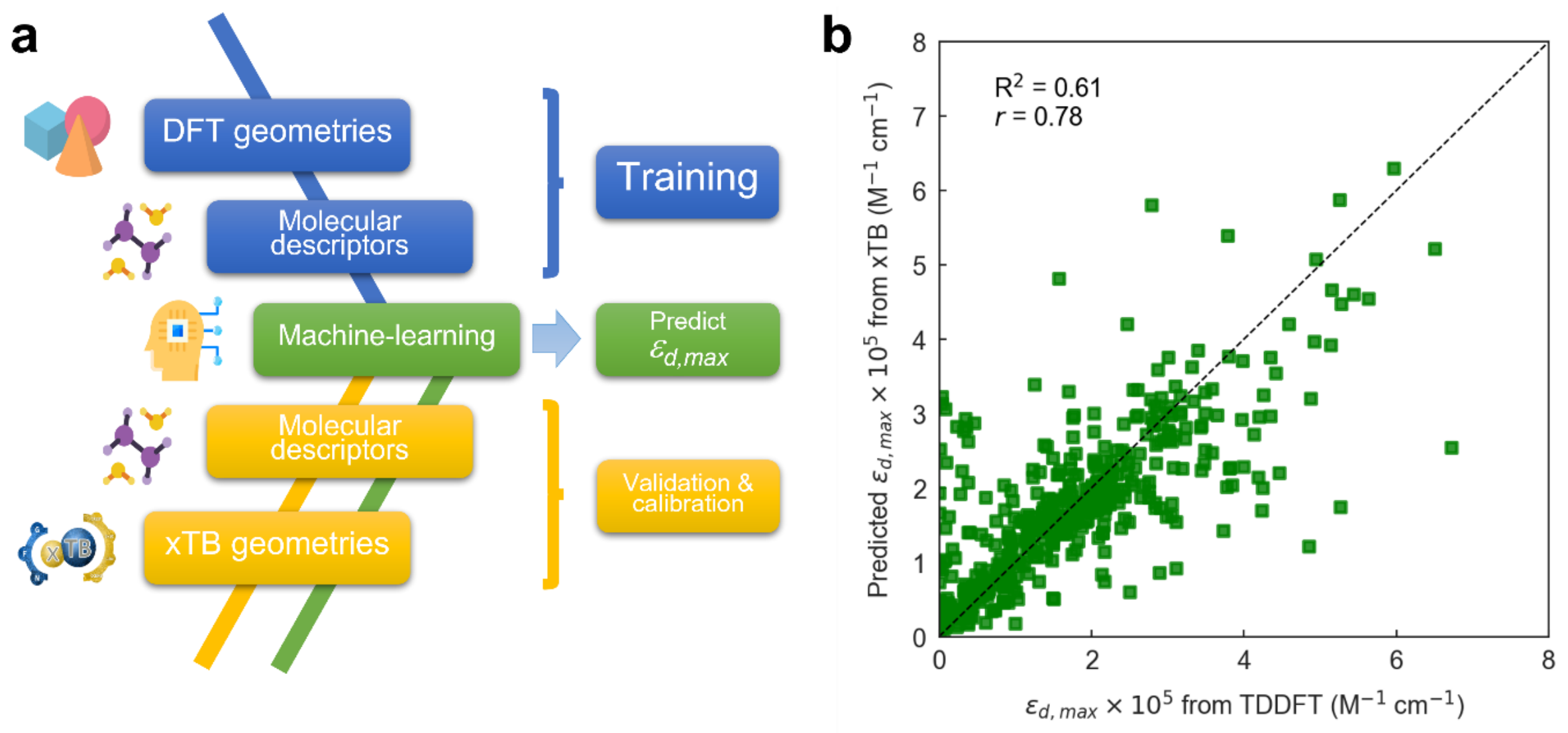

**Figure 5.** (a) ML workflow used in this work to draw $\varepsilon_{d,max}$ predictions. A RF model is trained on TDDFT data and interpolated (validated) on xTB geometries, including also their corresponding molecular descriptors. To improve the accuracy of the model, energy levels obtained using the GFN2–xTB Hamiltonian require calibration with TDDFT values (Figure S22). (b) Leave-one-out interpolation of the resulting RF model using three input molecular descriptors (including calibrated energy levels) and a 64-bit Morgan fingerprint vector.

## 3. Conclusion

We have demonstrated that TDDFT calculations agree reasonably well with the experimental maximum molar extinction coefficient ($\varepsilon_{d,max}$) in solution state by exploiting a database of TDDFT-optimized small molecular acceptors (NFAs) and donor oligomers collected over the years. This finding supports further analysis of the molecular dataset to identify structure-absorption relationships by means of statistical and machine-learning (ML) methods. Through the exploration of molecular descriptors, we identify two features that are strongly correlated with $\varepsilon_{d,max}$, namely the linearity and planarity of the molecule in the direction of maximum atomic polarizability variance; and the number of $sp^2$-hybridized carbon atoms bonded to two other carbons included in the molecule. These further suggest design rules that a highly absorbing NFA should possess a fully conjugated, planar and linear molecular backbone with more polarisable heteroatoms. We further identify that moieties such as thieno[3,2-b]thiophene (TT), thiophene (T), 2-(5,6-difluoro-3-oxo-2,3-dihydro-1H-inden-1-ylidene)malononitrile (2FIC), 2-(3-oxo-2,3-dihydro-1H-inden-1-ylidene)malononitrile (IC) and indaceno[1,2-b:5,6-b′]dithiophene (IDT) appear more frequently in molecules with the highest absorption strength. Finally, we demonstrate the feasibility of random decision forests (RFs) trained with a few (3) molecular descriptors and 64-bit Morgan fingerprint vectors to predict $\varepsilon_{d,max}$ in molecular geometries optimized by a computationally less demanding method such as extended tight-binding (xTB). This approach shows the ability to bypass thorough TDDFT calculations, thus facilitating high-throughput screening of absorption strength in conjugated molecules in combination with generative molecular models.

## 4. Experimental and theoretical methods

Excited state calculation database and experimental $\varepsilon_{\mathrm{d,max}}$ database: TDDFT results in this study are based on the functional B3LYP and were performed by present and past group members in Prof. Jenny Nelson's group at Imperial College London, making up more than 3500 entries (corresponding to 479 unique molecules). The majority of experimental solid state thin film $\varepsilon_{\mathrm{d,max}}$ values for NFAs shown in **Figure 1**a, b and c were measured using variable-angle spectroscopic ellipsometry (VASE) for the present study. Neat films were deposited from solution by either spin- or blade-coating on glass substrates at distinct thicknesses (typically ranging from 30 to 150 nm). Ellipsometry data were acquired at three to five angles of incidence (55°-75°) using a Sopralab GES-5E rotating polarizer spectroscopic ellipsometer (SEMILAB) coupled to a charge-coupled device (CCD) detector. Experimental solution $\varepsilon_{\mathrm{d,max}}$ were mostly collected from literature with a majority of data taken from Ref. [95], and Y5, Y6, and Y7 measured using UV-visible spectroscopy. The complete database and sources are presented in the Supplementary Database.

Theoretical description of molar extinction coefficient ($\varepsilon_d$): To calculate the molar extinction coefficient $\varepsilon_d$, let us start with defining the absorption coefficient $\alpha$ in a quantum picture (we stay with SI units for the moment). The absorption coefficient for transition from state 1 to state 2 can be defined as [20,96]

$$\frac{dI}{dx} = -\alpha_{12} I, \qquad \text{(Equation S1)}$$

where $I$ is light intensity, determined by the energy density of an electromagnetic wave via

$$I = \frac{1}{2} n c \epsilon_0 |E_0|^2, \qquad \text{(Equation S2)}$$

where $n$ is the refractive index, $\epsilon_0$ vacuum permittivity, $c$ the speed of light, and $E_0$ the amplitude of the electric field. For an electromagnetic wave, the rate of intensity attenuation $\frac{dI}{dx}$ is equal to the rate of loss of energy density from the field $-\frac{dU}{dt}$, and the latter is the product of transition rate $\Gamma_{12}$ and transition energy $\hbar\omega_{12}$, and we have

$$\frac{dI}{dx} = -N\Gamma_{12}\hbar\omega_{12}, \qquad \text{(Equation S3)}$$

where $N$ is the volume density of molecules and $\hbar$ the reduced Planck constant. Substituting for $\frac{dI}{dx}$ and I in the definition of $\alpha_{12}$ we get

$$\alpha_{12} = \frac{2N\hbar\omega_{12}\Gamma_{12}}{nc\epsilon_0|E_0|^2} \qquad \text{(Equation S4)}$$

The transition rate $\Gamma_{12}$ can be defined by Fermi's Golden Rule and the perturbing Hamiltonian given by $H = d_{12}\boldsymbol{E_0}$ using dipole approximation, where $d_{12}$ is the transition dipole moment of the transition. Considering randomly oriented transition dipoles relative to the direction of the exciting electromagnetic field, we have

$$\Gamma_{12} = \frac{2\pi}{3\hbar} d_{12}^2|E_0|^2\delta(\hbar\omega - E_2 + E_1) \qquad \text{(Equation S5)}$$

Using $E_2 - E_1 = \hbar\omega_{12}$, we get

$$\alpha_{12} = \frac{4\pi N\omega_{12}}{3nc\epsilon_0\hbar} d_{12}^2\delta(\omega - \omega_{12}) \qquad \text{(Equation S6)}$$

From an arbitrary transition from state i to state j, we can express above equation using oscillator strength of the transition ($f_{ij}$):

$$\alpha_{ij} = \frac{2\pi Ne^2}{3\epsilon_0 m_0 nc} f_{ij}\delta(\omega - \omega_{ij}), \qquad \text{(Equation S7)}$$

where $e$ is the elementary charge, and $f_{ij} = \frac{2m_0\omega_{12}}{e^2\hbar} d_{ij}^2$. Integrating over all transitions, we have

$$\alpha(\omega) = \frac{2\pi Ne^2}{3\epsilon_0 m_0 nc} \sum_{ij} f_{ij}\delta(\omega - \omega_{ij}) \qquad \text{(Equation S8)}$$

To correlate the absorption coefficient ($\alpha$) with the molar extinction coefficient ($\varepsilon_d$), we need the definition of optical density ($OD$) and optical depth ($\alpha d$). Light is attenuated by passing through a depth $d$ of material such that

$$I(d) = I_0 e^{-\alpha d} = I_0 10^{-OD} \qquad \text{(Equation S9)}$$

And optical density, or called sometimes absorbance is defined as

$$OD = \rho \varepsilon_d d, \qquad \text{(Equation S10)}$$

where $\rho$ is concentration in molar (M or mol $L^{-1}$), and $d$ is sample length in cm. Consequently, we have

$$\varepsilon_d = \frac{log_{10}(e)}{\rho} \alpha_{cm} = \frac{\alpha_{cm}}{2.303\rho}, \qquad \text{(Equation S11)}$$

noting that we now write the absorption coefficient per cm to distinguish from the expression for $\alpha$ above, which we did assuming SI units, hence $\alpha_{cm} = \frac{\alpha}{100}$. $\rho$ is moles of molecules per $dm^3$. We now have

$$\varepsilon_d(\omega) = 10\, log_{10}(e)\, N_A \frac{2\pi e^2}{3\epsilon_0 m_0 nc} \sum_{ij} f_{ij} \delta(\omega - \omega_{ij}) \qquad \text{(Equation S12)}$$

Let us recast this in terms of photon energy $E$ in eV, i.e. $E = \frac{\hbar\omega}{e}$, rather than angular frequency, so it is easier to consider the magnitude, and finally we have $\varepsilon_d$ in the unit of $M^{-1}$ $cm^{-1}$.

$$\varepsilon_d(E) = 10\, log_{10}(e)\, N_A \frac{2\pi e\hbar}{3\epsilon_0 m_0 nc} \sum_{ij} f_{ij} \delta(E - E_{ij}) \qquad \text{(Equation S13)}$$

This allows us to compute the theoretical $\varepsilon_d$ using the calculated oscillator strength at different transitions. And the common method to calculate the oscillator strength is time-dependent-density-functional-theory (aka TDDFT).

Converting complex refractive index from solid state ellipsometry measurements to $\varepsilon_d$: Using ellipsometry measurements from film (solid state), we can extract the complex refractive index, $\eta$

$$\eta = n + i\kappa \qquad \text{(Equation S14)}$$

Where $n$ is the refractive index, and $\kappa$ the extinction coefficient. The absorption coefficient ($\alpha_{cm}$) is then determined by

$$\alpha_{cm} = \frac{4\pi\kappa}{\lambda_{cm}} \qquad \text{(Equation S15)}$$

Where $\lambda_{cm}$ is the wavelength in centimetre. Using Equation S11, and the relationship between molar concentration $\rho$ and mass concentration $\rho_M$, i.e., $\rho = \frac{\rho_M}{M_w}$, we have

$$\varepsilon_d = log_{10}(e)\,\alpha_{cm}\frac{M_w}{\rho_M} = log_{10}(e)\frac{4\pi\kappa}{\lambda_{cm}}\frac{M_w}{\rho_M} \qquad \text{(Equation S16)}$$

Where $M_w$ is the molecular weight, g $mol^{-1}$, and $\rho_M$ has the unit of g $L^{-1}$, and is typically assumed to be 1000 g $L^{-1}$.

**Author contributions**

J.Y. and X.R.-M. contributed equally to this work and drafted the paper. J.Y. performed DFT and TDDFT calculations, absorption strength analysis, and data collection. X.R.-M. performed the statistical analysis and machine-learning study. D.P., H.D., D.B., M.A., A.V., S.F., A.A.S., and X.H. shared their DFT/TDDFT calculation results. F.E. prepared thin films of NFAs for VASE measurements. X.R.-M., V.B., and B.D. did VASE measurements. E.R. did UV-vis measurements of Y5, Y6, and Y7 in solution. G.Z. and H.-L.Y. provided Y5, Y6, and Y7. All authors gave critical review on this work. J.N. and M.C.-Q. supervised this work.

**Conflicts of interest**

There are no conflicts to declare.

**Acknowledgements**

J.N., J.Y., D.P., M.A., F.E., and E.R. thank the European Research Council for support under the European Union's Horizon 2020 research and innovation program (Grant Agreement No. 742708 and No. 648901). The authors at ICMAB acknowledge financial support from the Spanish Ministry of Science and Innovation through the Severo Ochoa" Program for Centers of Excellence in R&D (No. CEX2019-000917-S), and project PGC2018-095411-B-I00. E.R. is grateful to the Fonds de Recherche du Quebec-Nature et technologies (FRQNT) for a postdoctoral fellowship and acknowledges financial support from the European Cooperation in Science and Technology. M.A. thanks the Engineering and Physical Sciences Research Council (EPSRC) for support via doctoral studentships. F.E. thanks the Engineering and Physical Sciences Research Council (EPSRC) for support via the Post-Doctoral Prize Fellowship. X.R.-M. acknowledges Prof. Olle Inganäs and the Knut and Allice Wallenberg Foundation for funding of his current postdoctoral position. H.-L. Yip thanks the support from Guangdong Major Project of Basic and Applied Basic Research (2019B030302007). The TOC figure and Figure 5a in the manuscript include freely available resources from Flaticon.com. J.Y. thank Xiaodan Ge for her support.

# Supporting Information

**Identifying structure-absorption relationships and predicting absorption strength of non-fullerene acceptors for organic photovoltaics**

*Jun Yan,[a,#] Xabier Rodríguez-Martínez,*[b,c,#] Drew Pearce,[a] Hana Douglas,[a] Danai Bili,[a] Mohammed Azzouzi,[a] Flurin Eisner,[a] Alise Virbule,[a] Elham Rezasoltani,[a] Valentina Belova,[c] Bernhard Dörling,[c] Sheridan Few,[a,f] Anna A. Szumska,[a] Xueyan Hou,[a] Guichuan Zhang,[d] Hin-Lap Yip,[d,e] Mariano Campoy-Quiles *[c] and Jenny Nelson*[a]*

[#] J.Y. and X.R.-M. contributed equally to this work.

**a** Department of Physics, Imperial College London, SW7 2AZ, London, United Kingdom
Email: jenny.nelson@imperial.ac.uk

**b** Electronic and Photonic Materials (EFM), Department of Physics, Chemistry and Biology (IFM), Linköping University, Linköping, SE 581 83 Sweden
Email: xabier.rodriguez.martinez@liu.se

**c** Instituto de Ciencia de Materiales de Barcelona, ICMAB-CSIC, Campus UAB, Bellaterra 08193, Spain
Email: mcampoy@icmab.es

**d** Institute of Polymer Optoelectronic Materials and Devices, State Key Laboratory of Luminescent Materials and Devices, South China University of Technology, Guangzhou 510640, P. R. China

**e** Department of Materials Science and Engineering, City University of Hong Kong, Tat Chee Avenue, Kowloon, Hong Kong

**f** Sustainability Research Institute, School of Earth and Environment, University of Leeds, Leeds, LS2 9JT

**Supplementary Note 1.** Chemical names and nomenclature of the materials highlighted in this work.

**PC61BM**: [6,6]-Phenyl-$C_{61}$-butyric acid methyl ester

**PC71BM**: [6,6]-Phenyl-$C_{71}$-butyric acid methyl ester

**ICBA**: 1′,1″,4′,4″-Tetrahydro-di[1,4]methanonaphthaleno[1,2:2′,3′,56,60:2″,3″][5,6]fullerene-$C_{60}$

**Y5**: (2,2'-((2Z,2'Z)-((12,13-bis(2-ethylhexyl)-3,9-diundecyl-12,13-dihydro[1,2,5]thiadiazolo[3,4e]thieno[2",3":4',5']thieno[2',3':4,5]pyrrolo[3,2-g] thieno[2',3':4,5]thieno[3,2-b]-indole-2,10-diyl)bis(methanylylidene))bis(3-oxo-2,3-dihydro-1H-indene-2,1-diylidene))dimalononitrile)

**Y6**: 2,2'-((2Z,2'Z)-((12,13-bis(2-ethylhexyl)-3,9-diundecyl-12,13-dihydro-[1,2,5]thiadiazolo[3,4-e]thieno[2",3'':4',5']thieno[2',3':4,5]pyrrolo[3,2-g]thieno[2',3':4,5]thieno[3,2-b]indole-2,10-diyl)bis(methanylylidene))bis(5,6-difluoro-3-oxo-2,3-dihydro-1H-indene-2,1-diylidene))dimalononitrile

**Y7**: 2,2'-((2Z,2'Z)-((12,13-bis(2-ethylhexyl)-3,9-diundecyl-12,13-dihydro-[1,2,5]thiadiazolo[3,4-e]thieno[2",3'':4',5']thieno[2',3':4,5]pyrrolo[3,2-g]thieno[2',3':4,5]thieno[3,2-b]indole-2,10-diyl)bis(methanylylidene))bis(5,6-dichloro-3-oxo-2,3-dihydro-1H-indene-2,1-diylidene))dimalononitrile

**Y11**: 2,2'-((2Z,2'Z)-((6,12,13-tris(2-ethylhexyl)-3,9-diundecyl-12,13-dihydro-6H-thieno[2",3":4',5']thieno[2',3':4,5]pyrrolo[3,2-g]thieno[2',3':4,5]thieno[3,2-b][1,2,3]triazolo[4,5-e]indole-2,10-diyl)bis(methanylylidene))bis(5,6-difluoro-3-oxo-2,3-dihydro-1H-indene-2,1-diylidene))dimalononitrile

**Y12**: 2,2'-((2Z,2'Z)-((12,13-bis(2-butyloctyl)-3,9-diundecyl-12,13-dihydro-[1,2,5]thiadiazolo[3,4-e]thieno[2",3'':4',5']thieno[2',3':4,5]pyrrolo[3,2-g]thieno[2',3':4,5]thieno[3,2-b]indole-2,10-diyl)bis(methanylylidene))bis(5,6-difluoro-3-oxo-2,3-dihydro-1H-indene-2,1-diylidene))dimalononitrile

**O-IDTBR**: (5Z,5'Z)-5,5'-(((4,4,9,9-tetrakis(n-octyl)-4,9-dihydro-s-indaceno[1,2-b:5,6-b']dithiophene-2,7-diyl)bis(benzo[c][1,2,5]thiadiazole-7,4-diyl))bis(methaneylylidene))bis(3-ethyl-2-thioxothiazolidin-4-one)

**O-IDFBR**: (5Z,5'Z)-5,5'-(((6,6,12,12-tetraoctyl-6,12-dihydroindeno[1,2-b]fluorene-2,8-diyl)bis(benzo[c][1,2,5]thiadiazole-7,4-diyl))bis(methaneylylidene))bis(3-ethyl-2-thioxothiazolidin-4-one)

**EH-IDTBR**: (5Z,5'Z)-5,5'-(((4,4,9,9-tetrakis(2-ethylhexyl)-4,9-dihydro-s-indaceno[1,2-b:5,6-b']dithiophene-2,7-diyl)bis(benzo[c][1,2,5]thiadiazole-7,4-diyl))bis(methaneylylidene))bis(3-ethyl-2-thioxothiazolidin-4-one)

**IDIC**: 2,2'-((2Z,2'Z)-((4,4,9,9-tetrahexyl-4,9-dihydro-s-indaceno[1,2-b:5,6-b']dithiophene-2,7-diyl)bis(methanylylidene))bis(3-oxo-2,3-dihydro-1H-indene-2,1-diylidene))dimalononitrile

**SN6IC-4F**: 2,2'-((2Z,2'Z)-((thieno[3,2-b]thieno[2''',3''':4'',5'']pyrrolo[2'',3'':4',5']thieno[2',3':4,5]thieno[2,3- d]pyrrole,4,9-dihydro-4,9-di-1-octylnonyl-2,7-diyl)bis(methanylylidene))bis((5,6-difluoro-3-oxo-2,3 -dihydro-1H-indene-2,1-diylidene))dimalononitrile

**ITIC**: 2,2'-[[6,6,12,12-tetrakis(4-hexylphenyl)-6,12-dihydrodithieno[2,3-d:2',3'-d']-s-indaceno[1,2-b:5,6-b']dithiophene-2,8-diyl]bis[methylidyne(3-oxo-1H-indene-2,1(3H)-diylidene)]]bis[propanedinitrile]

**ITIC-$C_2C_6$**: 2,2'-[[6,6,12,12-tetrakis(2-ethylhexyl)-6,12-dihydrodithieno[2,3-d:2',3'-d']-s-indaceno[1,2-b:5,6-b']dithiophene-2,8-diyl]bis[methylidyne(3-oxo-1H-indene-2,1(3H)-diylidene)]]bis[propanedinitrile]

**ITIC-$C_8$**: 2,2'-[[6,6,12,12-tetrakis(n-octyl)-6,12-dihydrodithieno[2,3-d:2',3'-d']-s-indaceno[1,2-b:5,6-b']dithiophene-2,8-diyl]bis[methylidyne(3-oxo-1H-indene-2,1(3H)-diylidene)]]bis[propanedinitrile]

**IT-4F**: 9-Bis(2-methylene-((3-(1,1-dicyanomethylene)-6,7-difluoro)-indanone))-5,5,11,11-tetrakis(4-hexylphenyl)-dithieno[2,3-d:2’,3’-d’]-s-indaceno[1,2-b:5,6-b’]dithiophene

**CBM**: 2,2'-(7,7'-(9-(heptadecan-9-yl)-9H-carbazole-2,7-diyl)bis(benzo[c][1,2,5]thiadiazole-7,4-diyl))bis(methan-1-yl-1-ylidene)dimalononitrile

**FBR**: 5,5′-[(9,9-Dioctyl-9H-fluorene-2,7-diyl)bis(2,1,3-benzothiadiazole-7,4-diylmethylidyne)]bis[3-ethyl-2-thioxo-4-thiazolidinone]

**BTPM**: 2,2'-((2Z,2'Z)-((12,13-diisobutyl-3,9-dimethyl-5,7,12,13-tetrahydro-[1,2,5]thiadiazolo[3,4-e]thieno[2",3":4',5']thieno[2',3':4,5]pyrrolo[3,2-g]thieno[2',3':4,5]thieno[3,2-b]indole-2,10-diyl)bis(methaneylylidene))bis(6-methyl-3-oxo-2,3-dihydro-1H-indene-2,1-diylidene))dimalononitrile

**BTTPC**: 2,2'-((6Z,6'Z)-((12,13-diisobutyl-3,9-dimethyl-5,7,12,13-tetrahydro-[1,2,5]thiadiazolo[3,4-e]thieno[2",3":4',5']thieno[2',3':4,5]pyrrolo[3,2-g]thieno[2',3':4,5]thieno[3,2-b]indole-2,10-diyl)bis(methaneylylidene))bis(5-oxo-5,6-dihydro-7H-indeno[5,6-b]thiophene-6,7-diylidene))dimalononitrile

**BTDTP-4F**: 2,2'-((2Z,2'Z)-((3,12-dimethyl-13,14-dihydro-12H-[1,2,5]thiadiazolo[3,4-e]thieno[2",3":4',5']pyrrolo[2',3':4,5]thieno[3,2-b]thieno[2",3":4',5']thieno[2',3':4,5]pyrrolo[3,2-g]indole-2,10-diyl)bis(methaneylylidene))bis(5,6-difluoro-3-oxo-2,3-dihydro-1H-indene-2,1-diylidene))dimalononitrile

**BDTP-4F**: 2,2'-(((1Z,1'Z)-(1,11-dimethyl-4,6,6c,10,11,11b,12,13-octahydro-2H-[1,2,5]thiadiazolo[3,4-e]thieno[2",3":4',5']pyrrolo[2',3':4,5]thieno[3,2-b]thieno[2',3':4,5]pyrrolo[3,2-g]indole-2,9(1H)-diylidene)bis(methaneylylidene))bis(5,6-difluoro-3-oxo-2,3,3a,6,7,7a-hexahydro-1H-indene-2-yl-1-ylidene))dimalononitrile

**BTTPTP-2OYPD**: 2,2'-((2Z,2'Z)-((13,14-diisobutyl-5,7,13,14-tetrahydro-[1,2,5]thiadiazolo[3,4-e]thieno[2''',3''':4",5"]thieno[2",3":4',5']thieno[2',3':4,5]pyrrolo[3,2-g]thieno[2",3":4',5']thieno[2',3':4,5]thieno[3,2-b]indole-2,10-diyl)bis(methaneylylidene))bis(3-oxo-2,3-dihydro-1H-indene-2,1-diylidene))dimalononitrile

**BTPTTT-2OYPD**: 2,2'-((2Z,2'Z)-((13,14-diisobutyl-5,7,13,14-tetrahydro-[1,2,5]thiadiazolo[3,4-e]thieno[2''',3''':4",5"]thieno[2",3":4',5']thieno[2',3':4,5]pyrrolo[3,2-g]thieno[2",3":4',5']thieno[2',3':4,5]thieno[3,2-b]indole-2,10-diyl)bis(methaneylylidene))bis(3-oxo-2,3-dihydro-1H-indene-2,1-diylidene))dimalononitrile

**IEICO**: 2,2′-((2Z,2′Z)-((5,5′-(4,4,9,9-tetrakis(4-hexylphenyl)-4,9-dihydros-indaceno[1,2-b:5,6-b′]dithiophene-2,7-diyl)bis(4-((2-ethylhexyl)oxy)thiophene-5,2-diyl))bis(methanylylidene))bis(3-oxo-2,3-dihydro-1H-indene-2,1-diylidene))dimalononitrile

**IEICO-4F**: 2,2'-((2Z,2'Z)-(((4,4,9,9-tetrakis(4-hexylphenyl)-4,9-dihydro-sindaceno[1,2-b:5,6-b']dithiophene-2,7-diyl)bis(4-((2-ethylhexyl)oxy)thiophene-5,2-

diyl))bis(methanylylidene))bis(5,6-difluoro-3-oxo-2,3-dihydro-1H-indene-2,1-diylidene))dimalononitrile

**BTTPTP-4Cl**: 2,2'-((2Z,2'Z)-((13,14-diisobutyl-5,7,13,14-tetrahydro-[1,2,5]thiadiazolo[3,4-e]thieno[2''',3''':4'',5'']thieno[2'',3'':4',5']thieno[2',3':4,5]pyrrolo[3,2-g]thieno[2'',3'':4',5']thieno[2',3':4,5]thieno[3,2-b]indole-2,10-diyl)bis(methaneylylidene))bis(5,6-dichloro-3-oxo-2,3-dihydro-1H-indene-2,1-diylidene))dimalononitrile

**INPIC-4F**: [(Z)-2-({24-[(Z)-(1-Dicyanomethylidene-5,6-difluoro-3-oxo-2-indanylidene)methyl]-15,15,30,30-tetrakis(p-hexylphenyl)-12,27-dioctyl-5,8,20,23-tetrathia-12,27-diazanonacyclo[16.12.0.$0^{3,16}$.$0^{4,14}$.$0^{6,13}$.$0^{7,11}$.$0^{19,29}$.$0^{21,28}$.$0^{22,26}$]triaconta-1(18),2,4(14),6(13),7(11),9,16,19(29),21(28),22(26),24-undecaen-9-yl}methylidene)-5,6-difluoro-3-oxo-1-indanylidene]propanedinitrile

**o-IO1**: 2-((Z)-2-((5-(7-(5-(((Z)-1-(dicyanomethylene)-5,6-difluoro-3-oxo-1,3-dihydro-2H-inden-2-ylidene)methyl)-3-((2-ethylhexyl)oxy)thiophen-2-yl)-4,4,9,9-tetraoctyl-4,9-dihydro-s-indaceno[1,2-b:5,6-b']dithiophen-2-yl)-4-(2-ethylhexyl)thiophen-2-yl)methylene)-5,6-difluoro-3-oxo-2,3-dihydro-1H-inden-1-ylidene)malononitrile

**TfIF-4FIC**: [(Z)-2-({26-[(Z)-(1-Dicyanomethylidene-5,6-difluoro-3-oxo-2-indanylidene)methyl]-7,7,16,16,23,23,32,32-octaoctyl-11,27-dithianonacyclo[17.13.0.$0^{3,17}$.$0^{4,15}$.$0^{6,13}$.$0^{8,12}$.$0^{20,31}$.$0^{22,29}$.$0^{24,28}$]dotriaconta-1(19),2,4(15),5,8(12),9,13,17,20(31),21,24(28),25,29-tridecaen-10-yl}methylidene)-5,6-difluoro-3-oxo-1-indanylidene]propanedinitrile

**NIBT**: (7Z,7'Z)-7,7'-(((4,4,9,9-tetrakis(4-octylphenyl)-4,9-dihydro-s-indaceno[1,2-b:5,6-b']dithiophene-2,7-diyl)bis(benzo[c][1,2,5]thiadiazole-7,4-diyl))bis(methaneylylidene))bis(2-(2-ethylhexyl)-1H-indeno[6,7,1-def]isoquinoline-1,3,6(2H,7H)-trione)

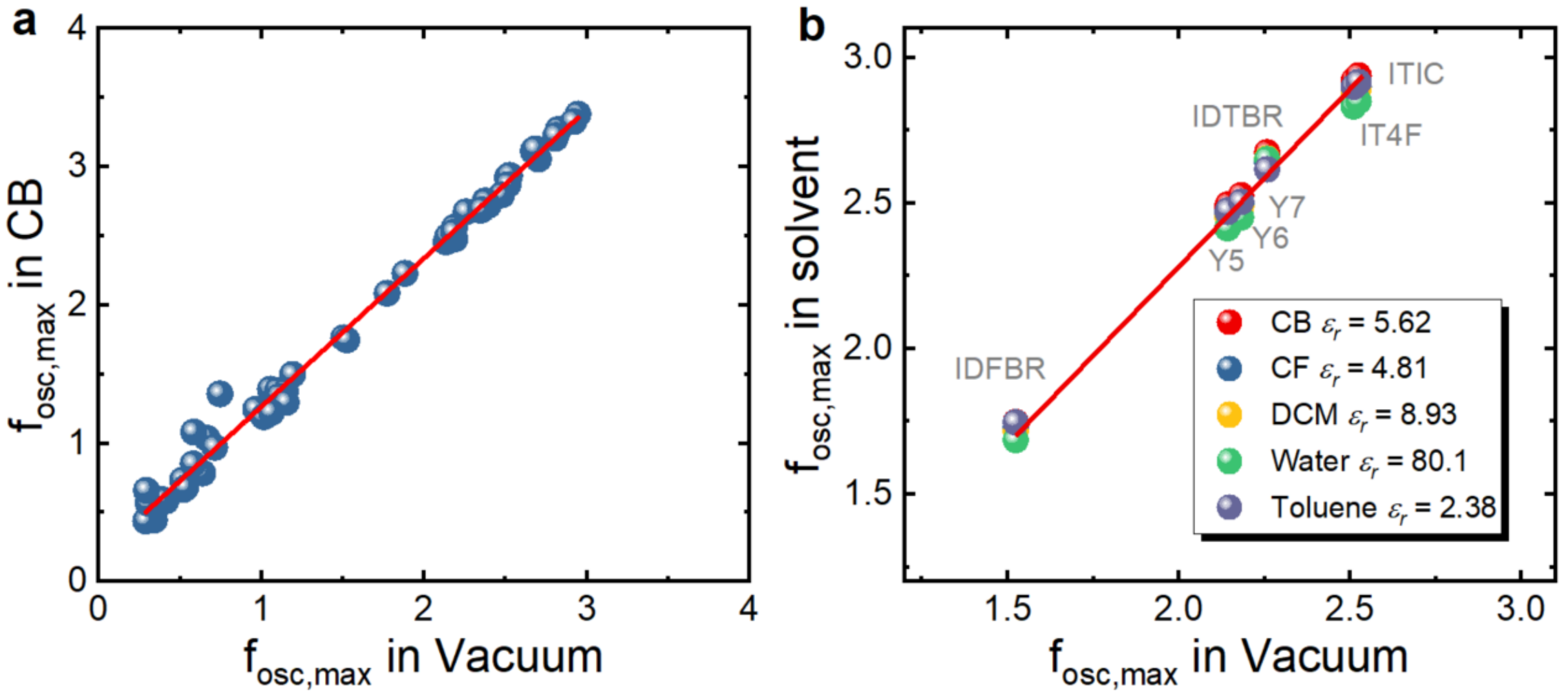

**Figure S1. Solvent effect on the maximum oscillator strength ($f_{osc,max}$) in TDDFT calculations. (a)** $f_{osc,max}$ in Chlorobenzene (CB) versus in vacuum for 56 organic molecules including common NFAs. **(b)** $f_{osc,max}$ in various organic solvents versus in vacuum for 7 different organic molecules, O-IDFBR, O-IDTBR, IT-4F, ITIC, Y5, Y6, and Y7. Noting here that $\varepsilon_{d,max} \propto f_{osc,max}$. TDDFT was performed under B3LYP/6-311+G(d,p) using Polarizable-continuum-solvent-model (PCM). We can see that the choice of solvent does not affect $f_{osc,max}$ much, and that a good linear correlation between solvent and vacuum $f_{osc,max}$ is obtained. This tells us that the same correlation between TDDFT and experiments will be maintained based on either vacuum environment or polarized medium, which allows us focus on TDDFT results from vacuum calculations only. This is a great benefit since most of the TDDFT calculations by the present and past group members were done in vacuum, allowing us to have a larger database.

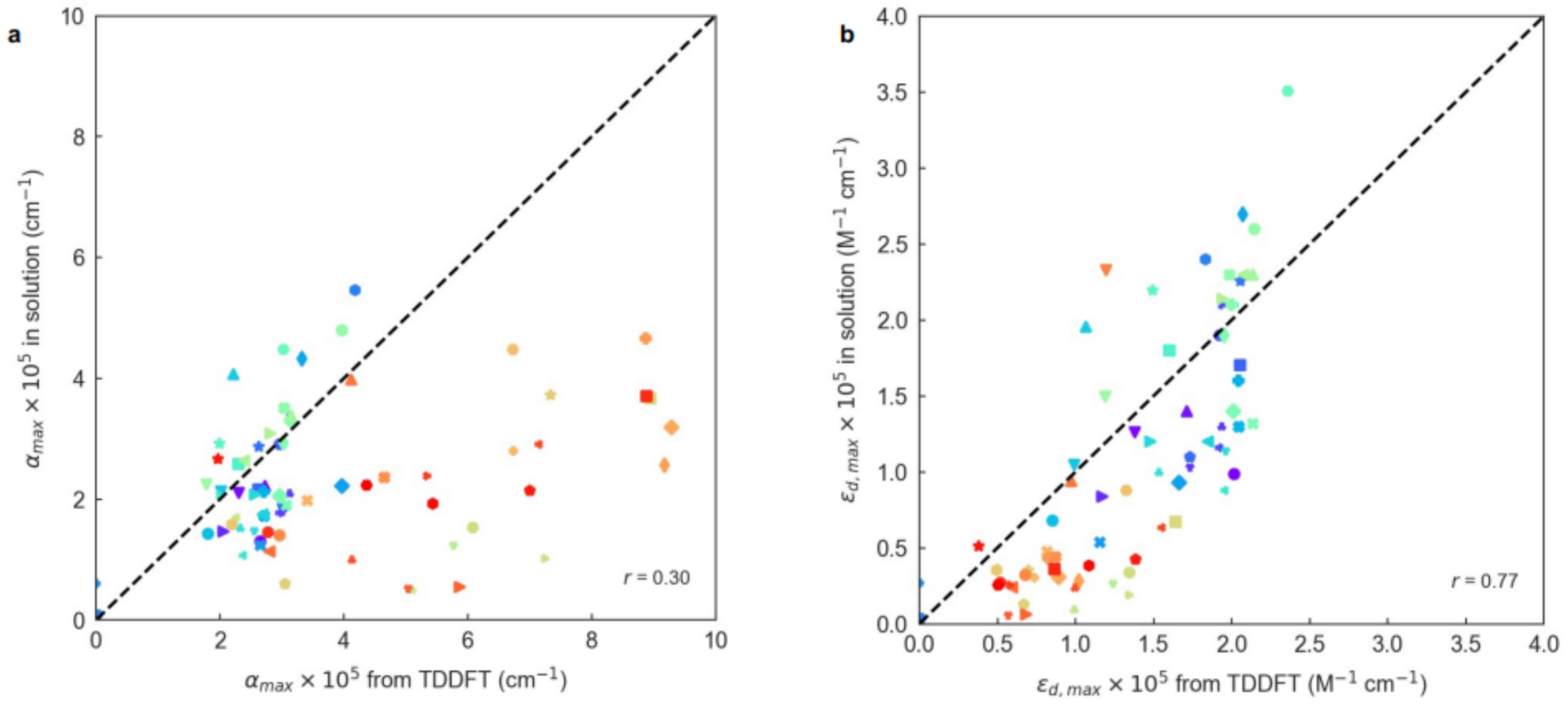

**Figure S2.** (a) Correlation between the maximum of the absorption coefficient ($\varepsilon_{d,max}$) obtained in solution state with the TDDFT calculated values. (b) Same dataset yet plotted in terms of maximum molar extinction coefficient $\varepsilon_{d,max}$). A significant fraction of this dataset was collected from literature.[95,97,106,107,98–105] When required, a refractive index of 1.5 and a solid density of 1000 g $L^{-1}$ were considered for all materials.

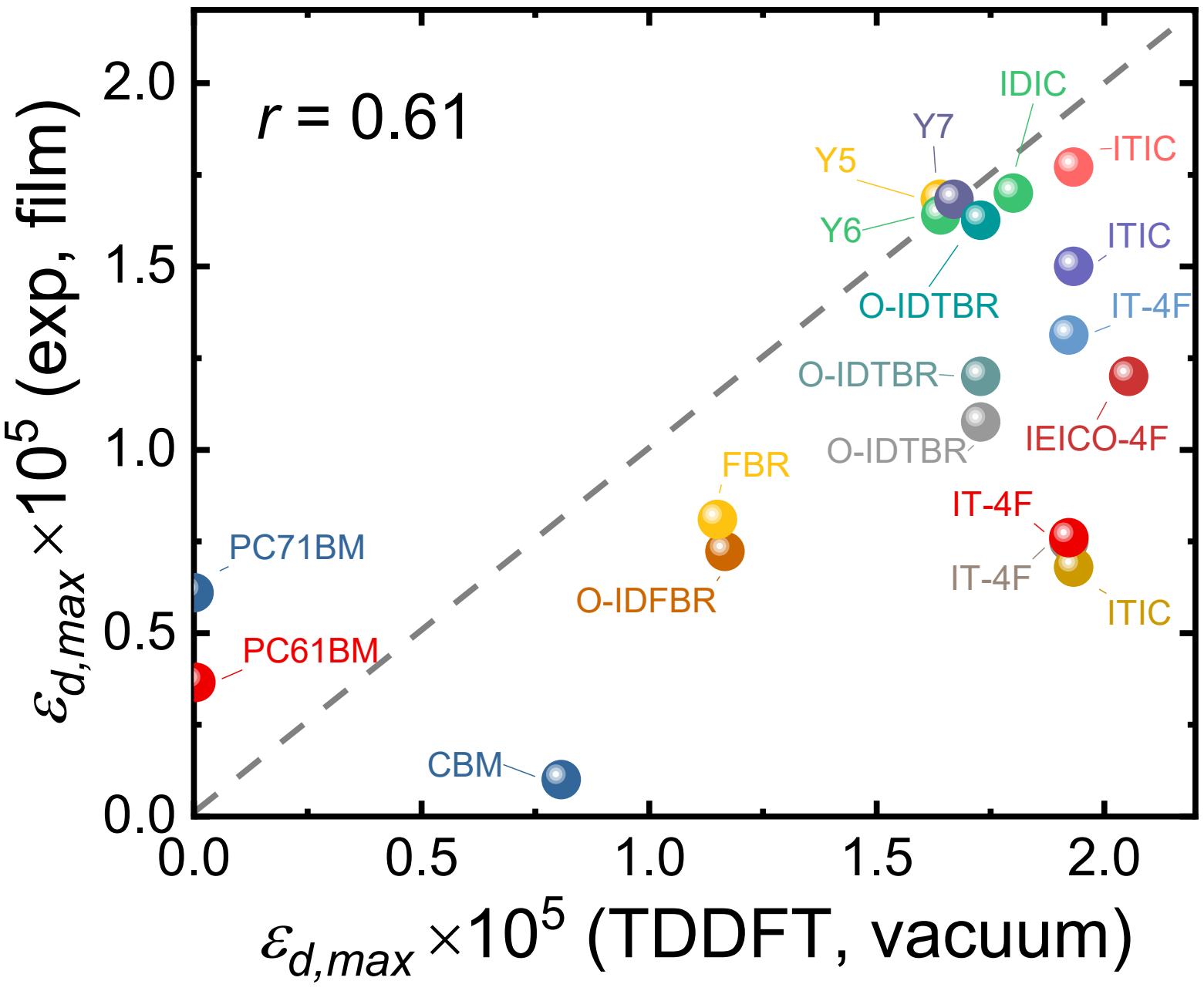

**Figure S3.** Comparison of maximum molar extinction coefficient ($\varepsilon_{d,max}$) between film experiments and TDDFT calculations in vacuum (20 data points). The experimental data of $\varepsilon_{d,max}$ in film are either measured or collected from literature, noting that different values may be present for the same material as they were collected from different papers. Unit of $\varepsilon_{d,max}$ is $M^{-1}$ $cm^{-1}$. Grey dashed lines indicate the perfect match between x- and y-axis. The detailed data for generating this figure is presented in the Supplementary Database.

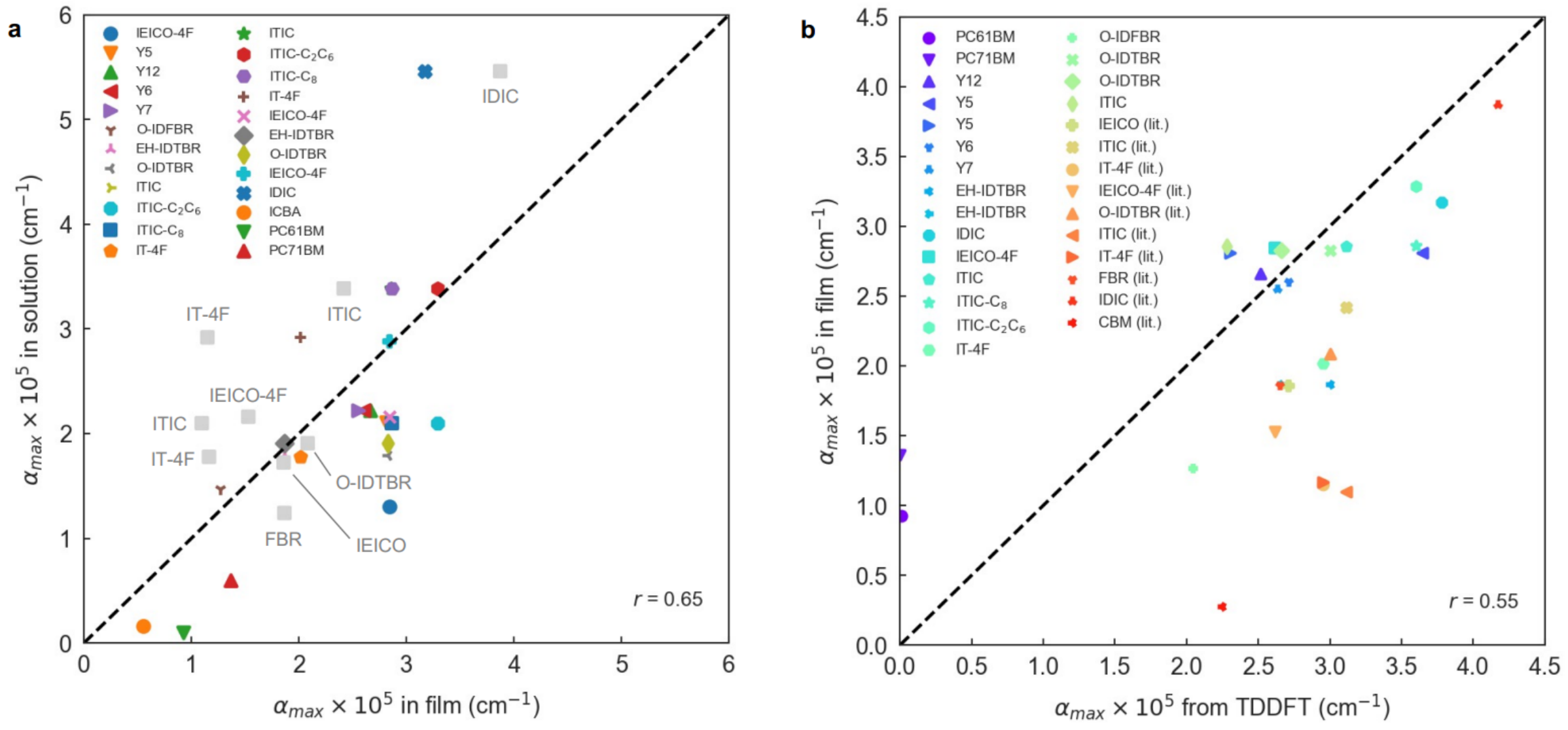

**Figure S4.** (a) Maximum absorption coefficient ($\alpha_{max}$) in solution and film. Grey squares correspond to data obtained from literature.[97,107] (b) Maximum absorption coefficient in film and as obtained in their corresponding TDDFT calculations. A few data points (labelled as lit.) correspond to values extracted from literature.[97,107]

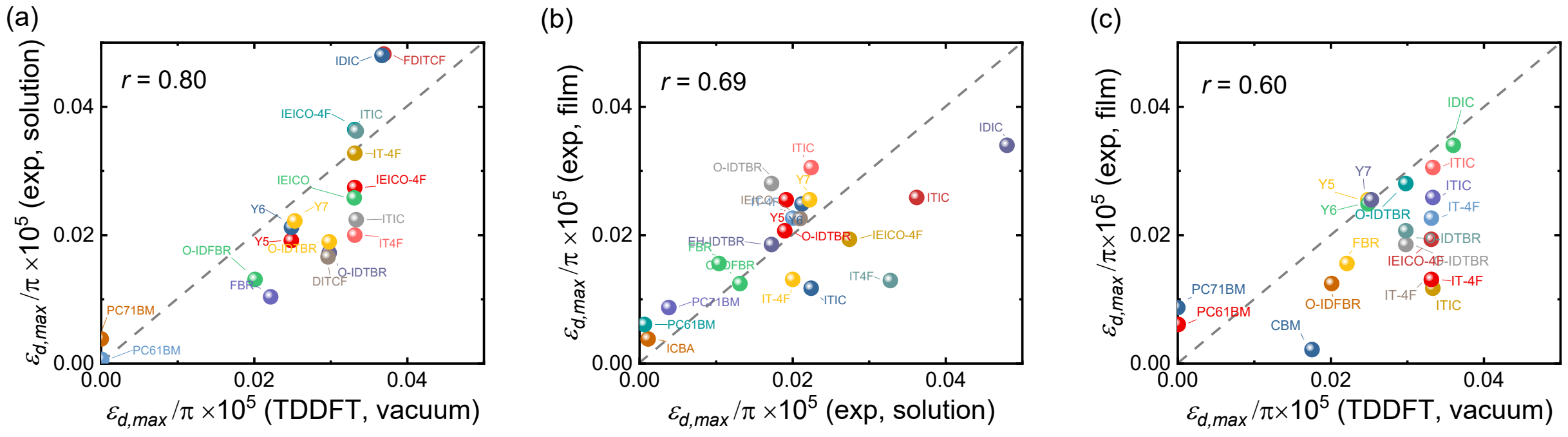

**Figure S5.** Effect of the number of π-electrons on the comparison between experimental (solution and film) maximum molar extinction coefficients and TDDFT results for the NFAs studied. (a) Experimental $\varepsilon_{d,max}/\pi$ in solution versus calculated $\varepsilon_{d,max}/\pi$ using TDDFT, (b) experimental $\varepsilon_{d,max}/\pi$ in film (solid state) versus that in solution; and (c) experimental $\varepsilon_{d,max}/\pi$ in solid state film versus calculated $\varepsilon_{d,max}/\pi$ using TDDFT. Unit of $\varepsilon_{d,max}/\pi$ is $M^{-1}$ $cm^{-1}$.

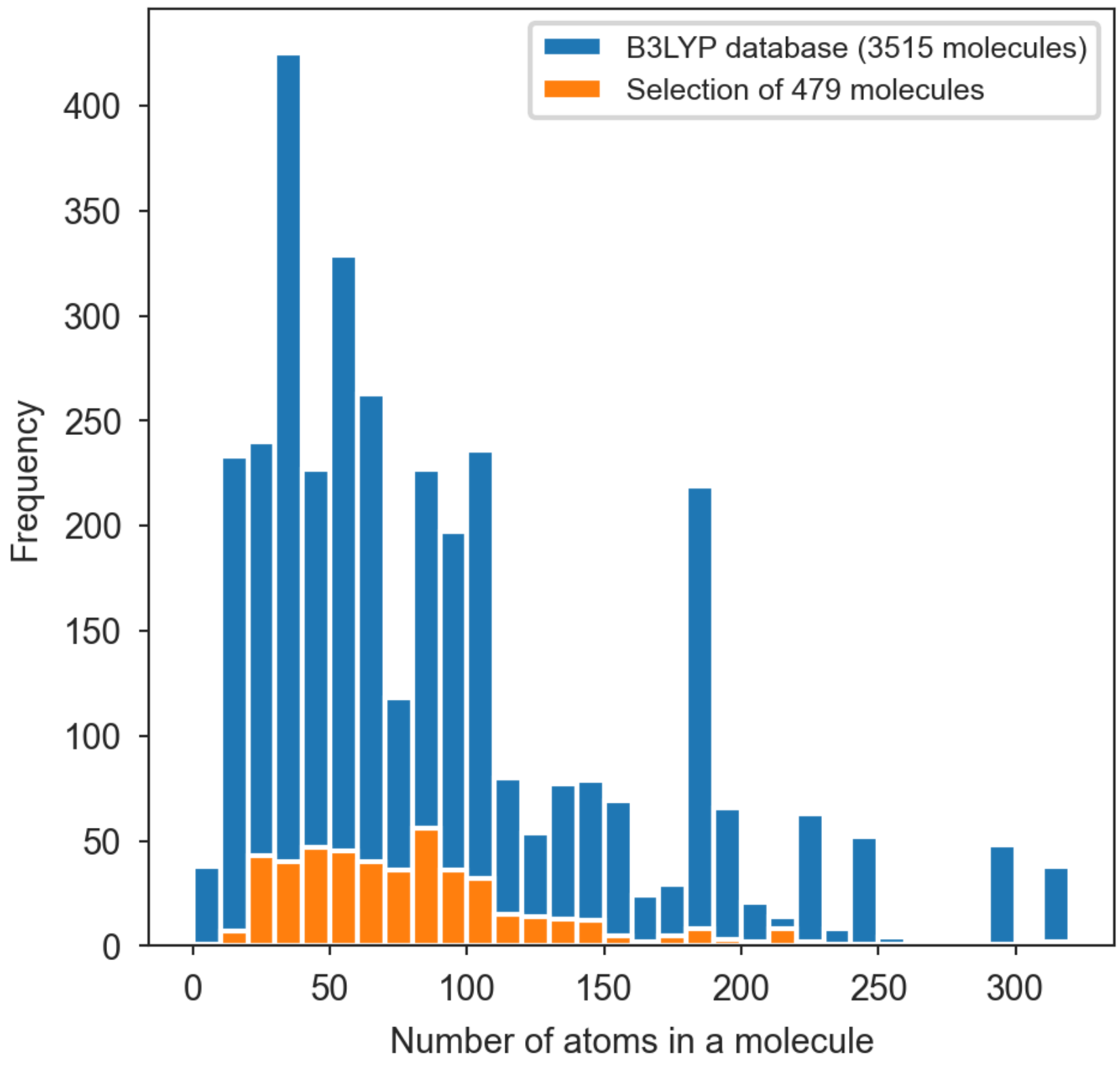

**Figure S6.** Histogram of the number of atoms present in the molecules of our TDDFT (B3LYP) dataset. Blue bars correspond to the molecules found originally in the dataset (3515 entries). Orange bars represent the distribution of the number of atoms found in the 479 molecules selected based on lowest energy conformation criteria.

**Supplementary Note 2.** Description of the TDDFT database, statistical and machine-learning methods.

The pristine data source of this work consists of a database of 3515 molecules optimized via DFT using the B3LYP functional as implemented in Gaussian09 software package. The database gathers original calculations performed for this particular study on conjugated small molecules as well as others developed in-house during the past years, including diverse conjugated small molecules, fullerenes and conjugated (co-)oligomers in distinct conformations (i.e., cis-/trans-).[31] Given the variety of input sources, the corresponding data cleaning procedure consists of: i) identifying duplicates based on molecular weight; and ii) picking the lowest energy molecular conformation among each set of duplicates. The filtering results in a final selection of 479 conjugated small molecules and oligomers optimized at the B3LYP level of theory. The resulting database gathers a variety of basis sets employed in the geometrical optimization step: 48% of the molecules were optimized using the 6-31G(d) set and 36% of them using the more computationally-expensive 6-311+G(d,p), see **Figure S7**. Furthermore, the chemical heterogeneity of the studied database is leaned toward known molecules and moieties of frequent use in high-performing solar energy harvesting applications, see Figure S8 and Figure S9. Side chains are systematically omitted or substituted by methyl groups in all calculations to reduce the computational cost.

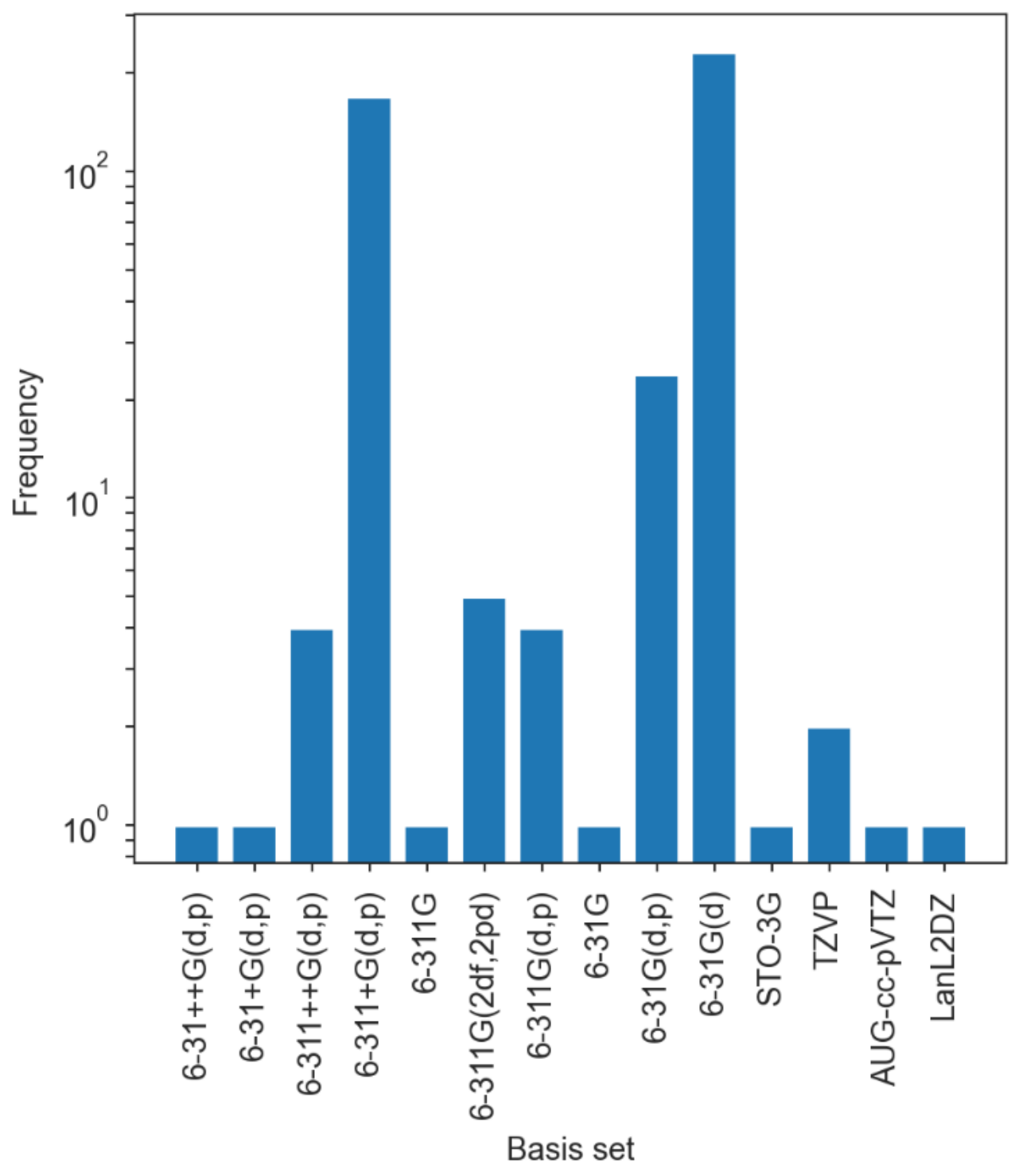

**Figure S7.** Histogram of the basis set employed in the geometrical optimization of the 479 molecules found in the curated DFT database.

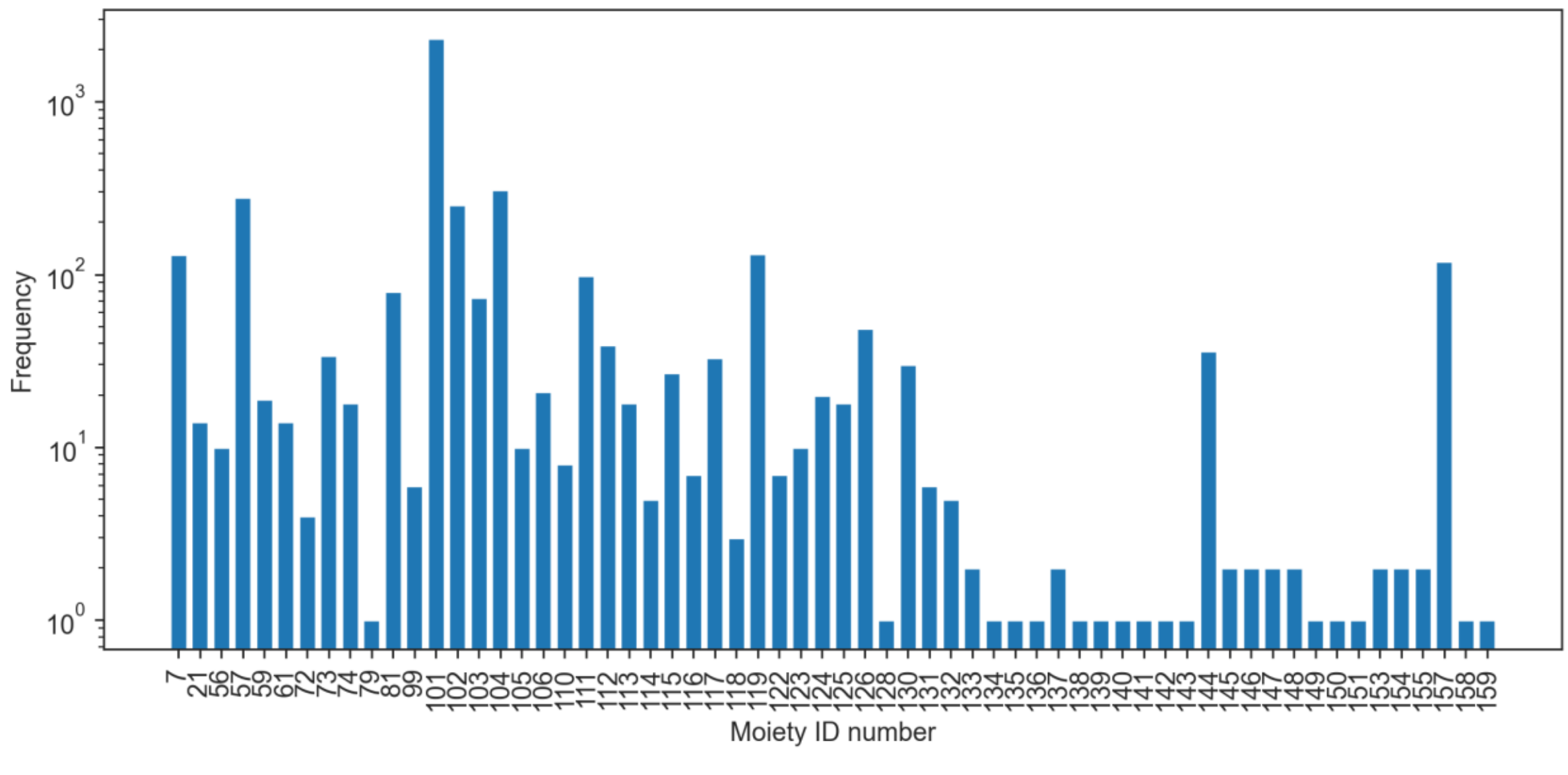

**Figure S8.** Histogram of the moieties present in the DFT database of 479 conjugated small molecules and oligomers. Moieties labels correspond to the chemical structures shown in Supplementary Note 4.

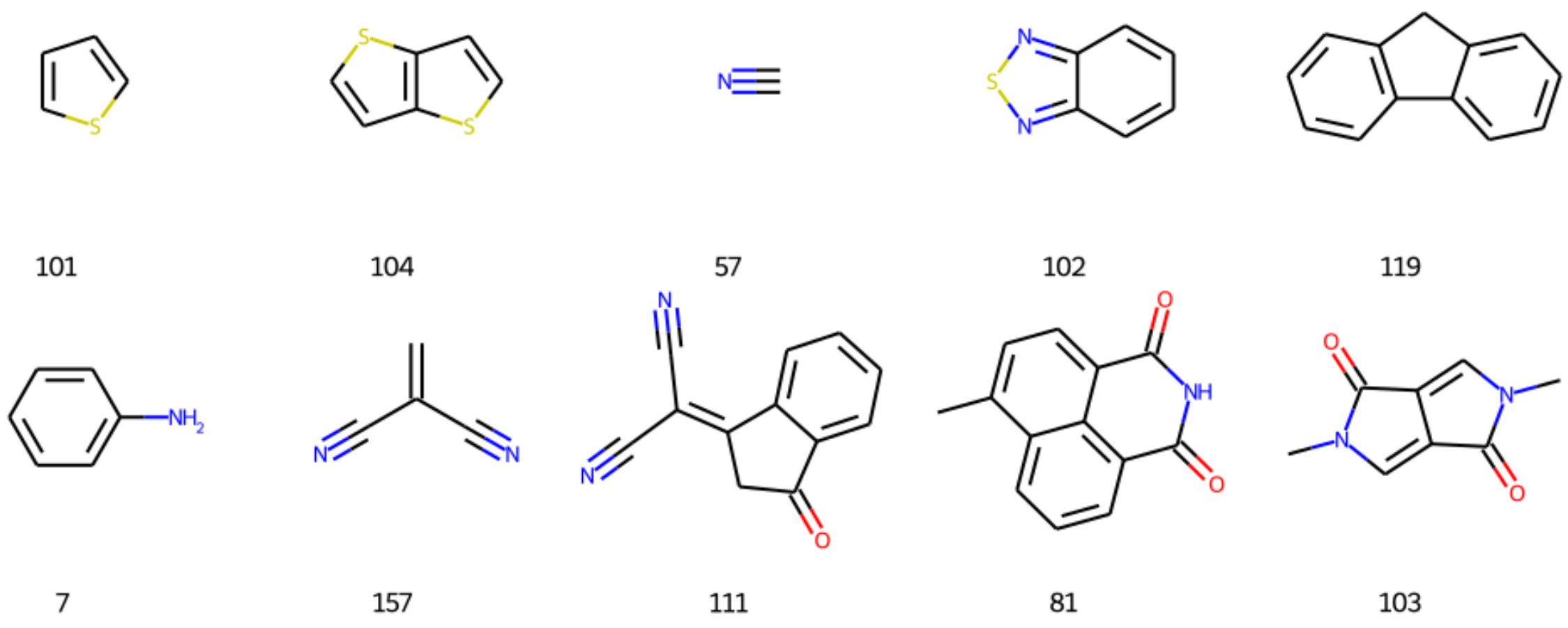

**Figure S9.** The 10 most frequent moieties together with their corresponding ID number.

Molecular descriptors are computed using four different open-source software packages and Python libraries (such as NumPy),[108] including 1D, 2D and 3D descriptors as retrieved from the corresponding DFT optimized geometries. The software bundle employed includes PaDEL (1874 descriptors),[75] PyChem (1094 descriptors),[76] Mordred (1826 descriptors)[77] and RDKit v2021.03.2. (1039 descriptors).[78] As a result, we obtained an initial set of 5834 descriptors for each molecule, which decreased up to 3239 after curing (i.e. dropping of uninformative or constant descriptors and others containing infinite or NaN values). Note that the same descriptor might be computed by more than one software bundle, yet slight numerical disagreements may arise due to the different computation algorithms. For that reason, we do not filter out redundant descriptors and perform their subsequent statistical analysis using the entire available catalogue. Furthermore, we include electronic features retrieved from the DFT calculations such as the energy levels of the 20 occupied and unoccupied molecular orbitals (HOMOs and LUMOs) closer to the band gap; the electronic band gap energy itself, $E_{gap}$; and the number of π electrons ($n_\pi$) in the molecule, which was determined using custom coding based on the RDKit library. The set of molecular fingerprints tested in this work is computed using RDKit and it includes customized coding for the moiety fingerprints and built-in functions for the computation of MACCS keys, Morgan fingerprints,[90] path-based (topological) fingerprints, E-state fingerprints[91] and Coulomb vectors.[92]

The target features in this study focus on the maximum oscillator strength ($f_{max}$) and other derived figures such as the maximum oscillator strength in the visible electromagnetic spectrum ($f_{max,vis}$, herein constrained between 300-1200 nm for its relevance in solar energy harvesting applications); the sum of $f$ in the visible window, $f_{max,vis}$; the spectral overlap between the Gaussian-broadened spectrum of fs in the visible (taking a standard deviation of 0.1 eV) and the AM1.5G solar irradiance spectrum, $f_{overlap}$; the maximum absorption coefficient ($\alpha_{max}$); the maximum of the imaginary part of the dielectric function ($\varepsilon_{2,max}$); and the maximum molar extinction coefficient, $f_{max}$, $f_{max,vis}$ and $f_{sum,vis}$ are also evaluated per number of π electrons in the molecule, i.e. $f_{max}/n_{\pi}$, $f_{max,vis}/n_{\pi}$ and $f_{sum,vis}/n_{\pi}$.

The statistical analysis of descriptors is deployed using the open-source library SciPy[109] whereas the machine-learning (ML) models (k-nearest neighbours, linear regression, support vector regressor and random forests) are implemented in Scikit-Learn.[89]

Regarding the scoring of the ML models, $R^2$ ranges from -∞ to unity, being 1 the best possible score and zero an indication of lack of predictive power (as it is always returning the expected value of the target function, i.e., its average value); $R^2_{adj}$ is formulated as[110]

$$R^2_{adj} = 1 - (1 - R^2)\frac{n-1}{n-p-1},$$

where $p$ is the number of variables and n the sample size. Thus, $R^2_{adj}$ adds penalties if the model uses too many variables, which is a useful metric when studying feature selection procedures. Test sets comprise 30% of the available data and all models are 10-fold holdout cross-validated (unless otherwise stated, using a randomized 70%-30% splitting for the train and test sets, respectively).

The recursive feature elimination (RFE) procedure applied in this work starts by decreasing the number of input features to 32 (i.e., 1% of the starting descriptor population of 3239 descriptors) in six consecutive feature reduction steps, in which after performing successive 10-fold cross-validations we drop 50% of the (averaged and least important) descriptors. Rather than observing a performance drop, the actual scoring of the RF ensemble improves as the number of features is reduced from 3239 ($R^2 = 0.65 \pm 0.06, r = 0.82 \pm 0.03$) to 51 ($R^2 = 0.70 \pm 0.05, r = 0.85 \pm 0.02$) variables in the last RFE iteration (**Figure S10**). After the last pruning step (51 variables), we select the 32 most important descriptors and perform a more thorough feature selection procedure by successively dropping (one-by-one) the least important descriptor (always keeping a 10-fold cross-validation scheme, see **Figure S11**).

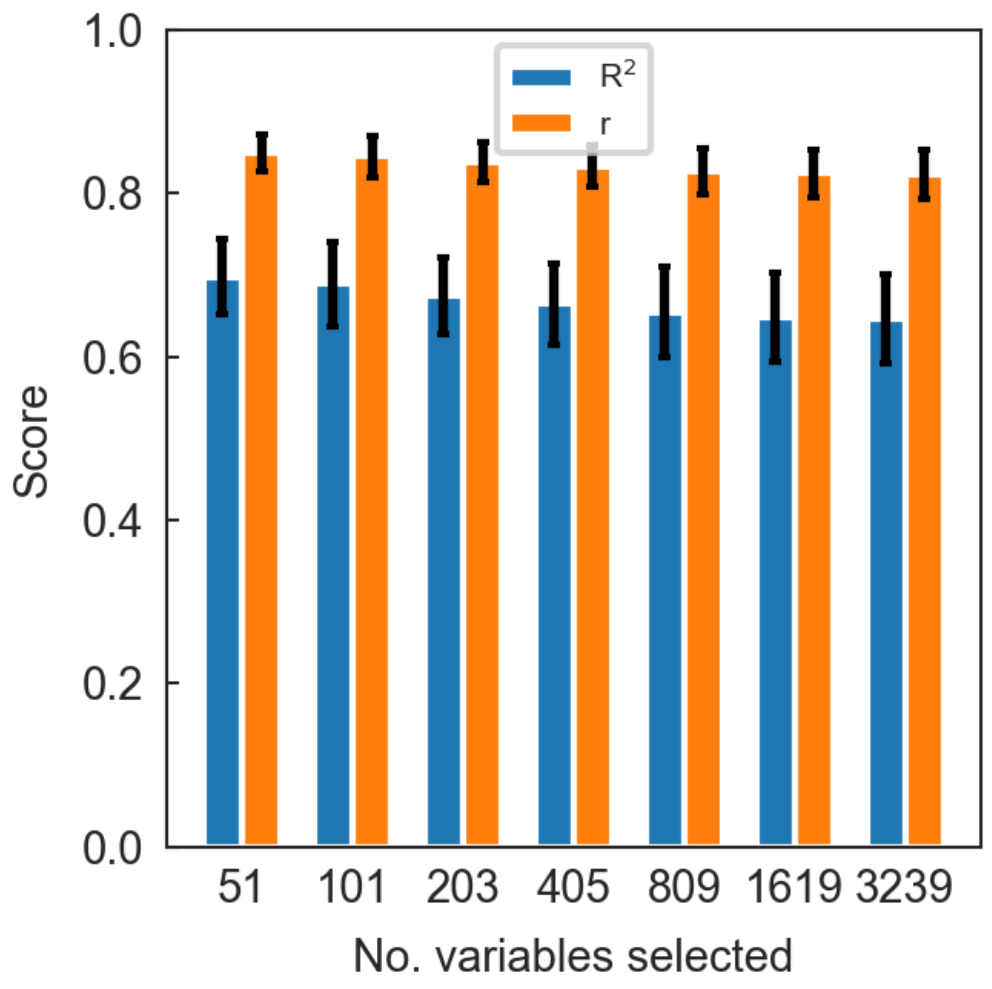

**Figure S10.** Scoring of RF regressors as part of a recursive feature elimination (RFE) loop in which 50% of the least important descriptors are dropped at each step.

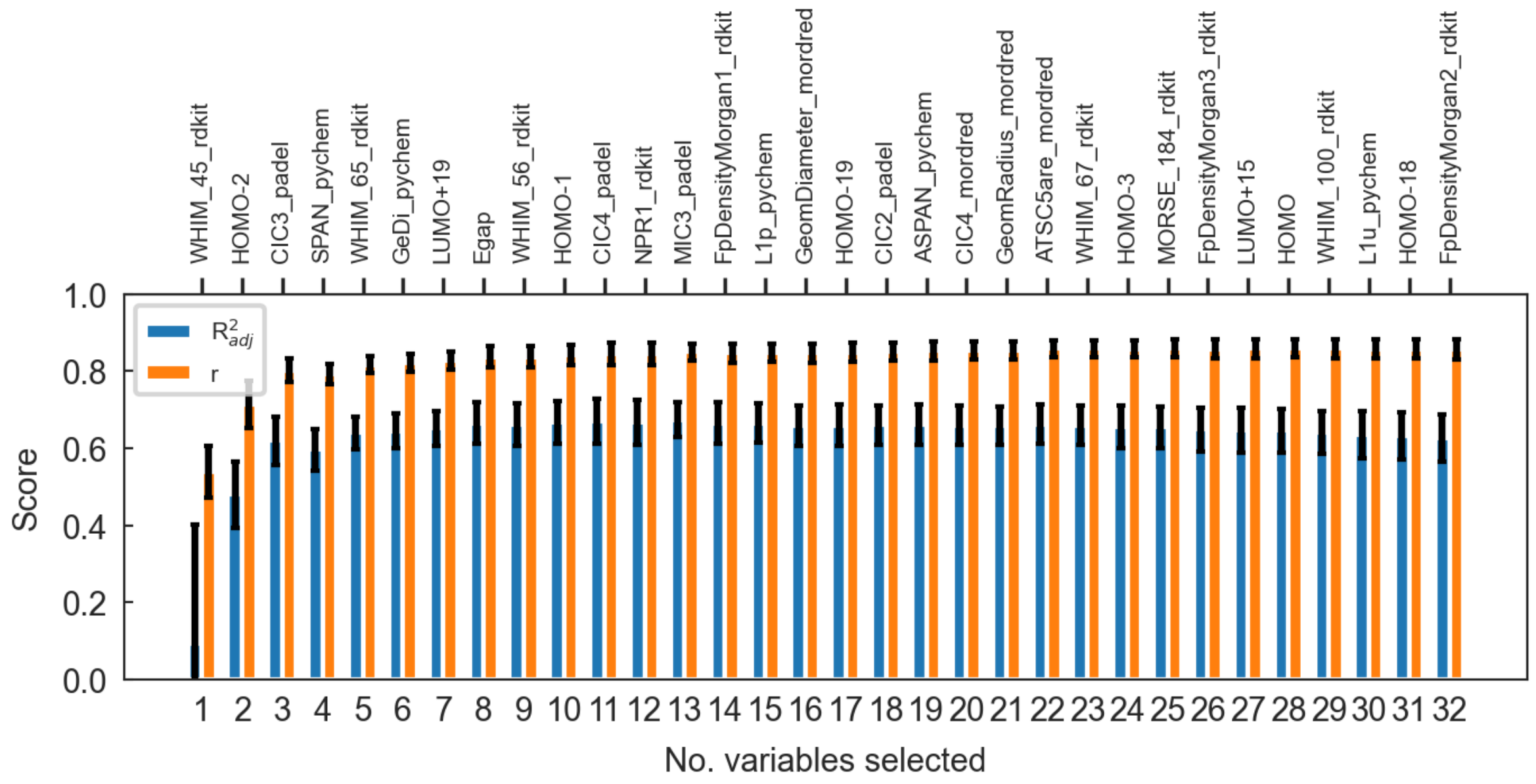

**Figure S11.** Scoring of 10-fold cross-validated RF regressors (300 estimators) trained and tested using different amounts of input descriptors as progressively indicated by the RFE algorithm. The top axis indicates, from left to right, the name of the variable that is added to the RF model, thus forming an ordered list of the most important descriptors found by the RF method.

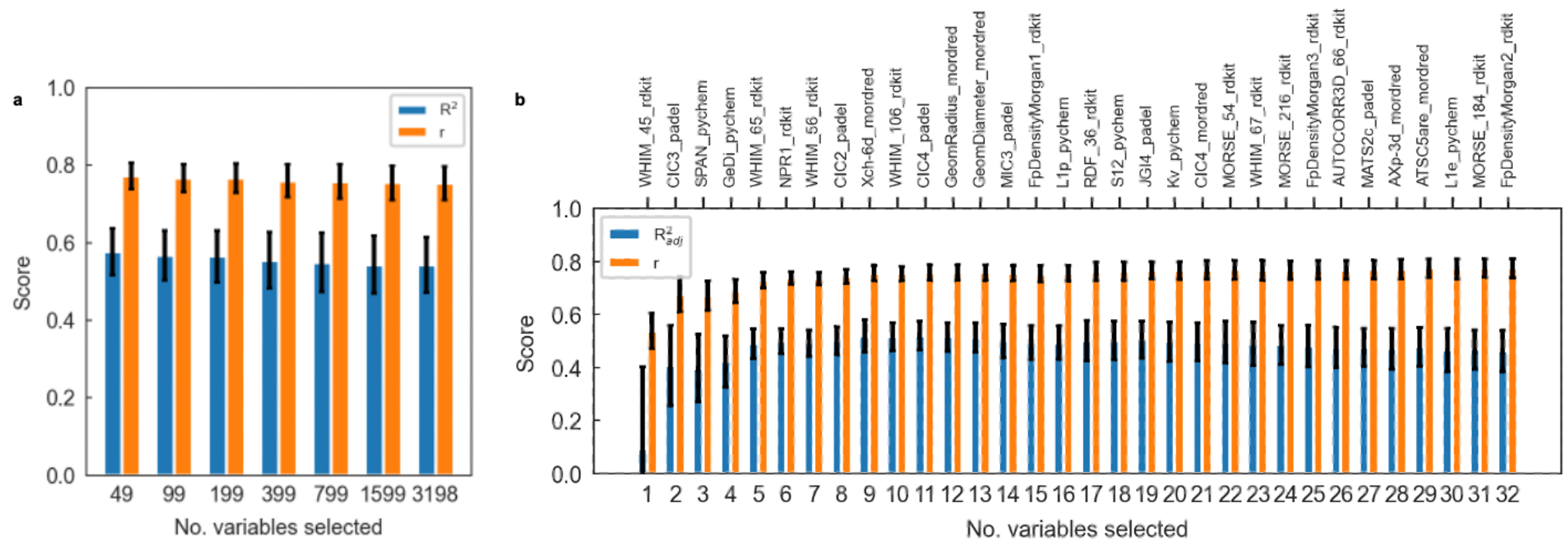

**Figure S12.** Performance of RF regressors trained without including electronic descriptors, using 300 estimators and 10-fold cross-validation. (a) Scoring parameters of cross-validated models as part of the RFE algorithm. (b) Detailed scoring parameters of the last 32 models obtained by RFE.

Supplementary Note 3. Clusterization algorithm of multicollinear descriptors.

The clusterization algorithm starts by taking the descriptor with the highest Spearman's rank correlation coefficient (ρ) and computing the Pearson correlation coefficient (r) with respect to the remaining elements in the $\rho$-ordered list of descriptors with $\rho \geq 0.7$. Descriptors from this list are dropped if $r \geq 0.7$ and considered to be in the same cluster; those showing $r \leq 0.7$ are candidates to form a different cluster. The process runs in a recursive-elimination manner until naturally leading to a selection of (typically) 1 to 5 descriptor clusters depending on the selected thresholds (0.6-0.7). These clusters gather the most statistically relevant and monotonic correlation trends with the target feature. Interestingly, by looking at the features stored in each of the clusters it is possible to replace some of the descriptors found originally by the algorithm by alternative figures of easier interpretation and/or larger physicochemical relevance.

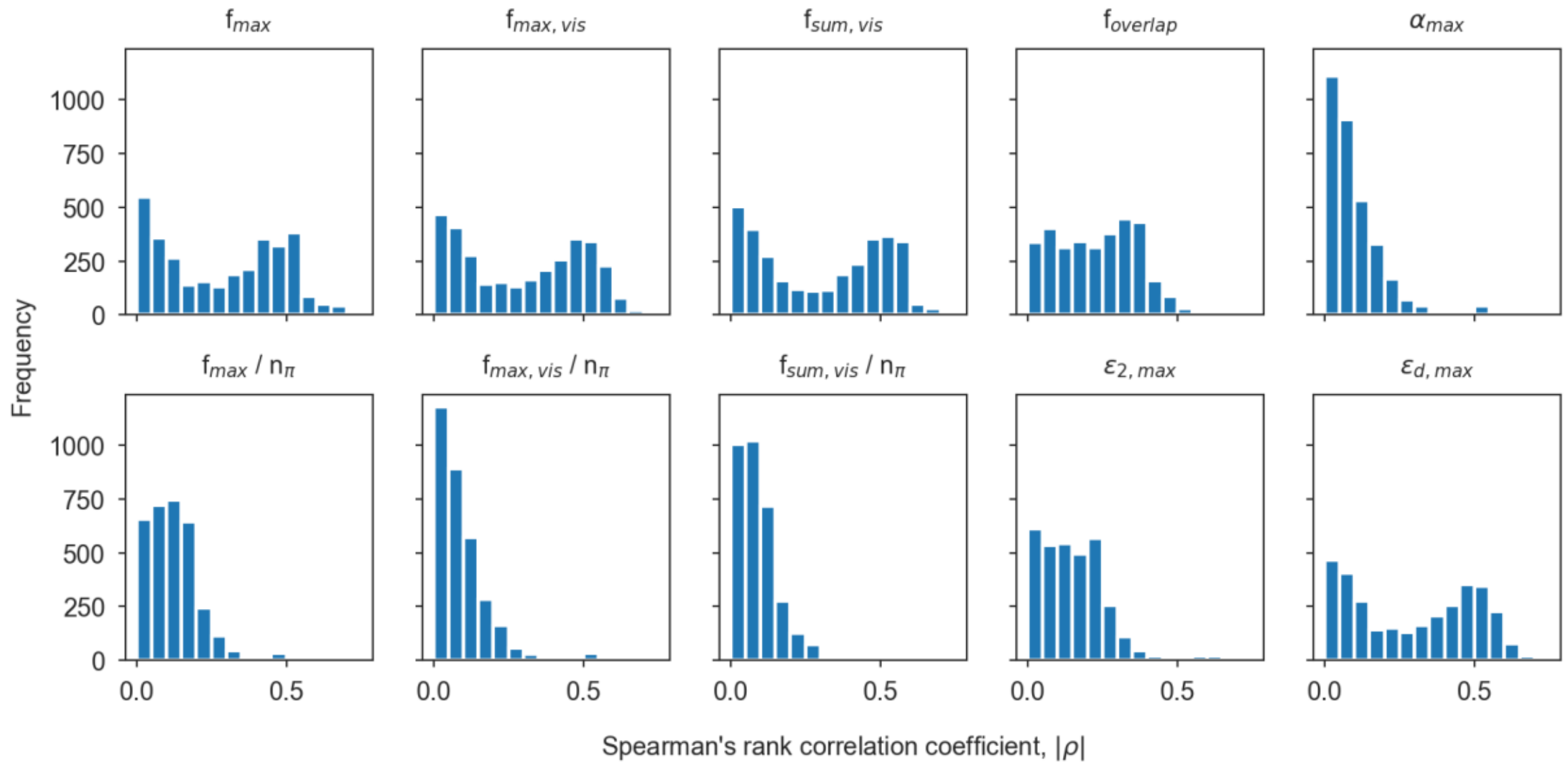

**Figure S13.** Spearman's rank correlation coefficient (in absolute value) histograms for the 3239 descriptors and the 10 different target features related with optical absorption and oscillator strength explored in this work.

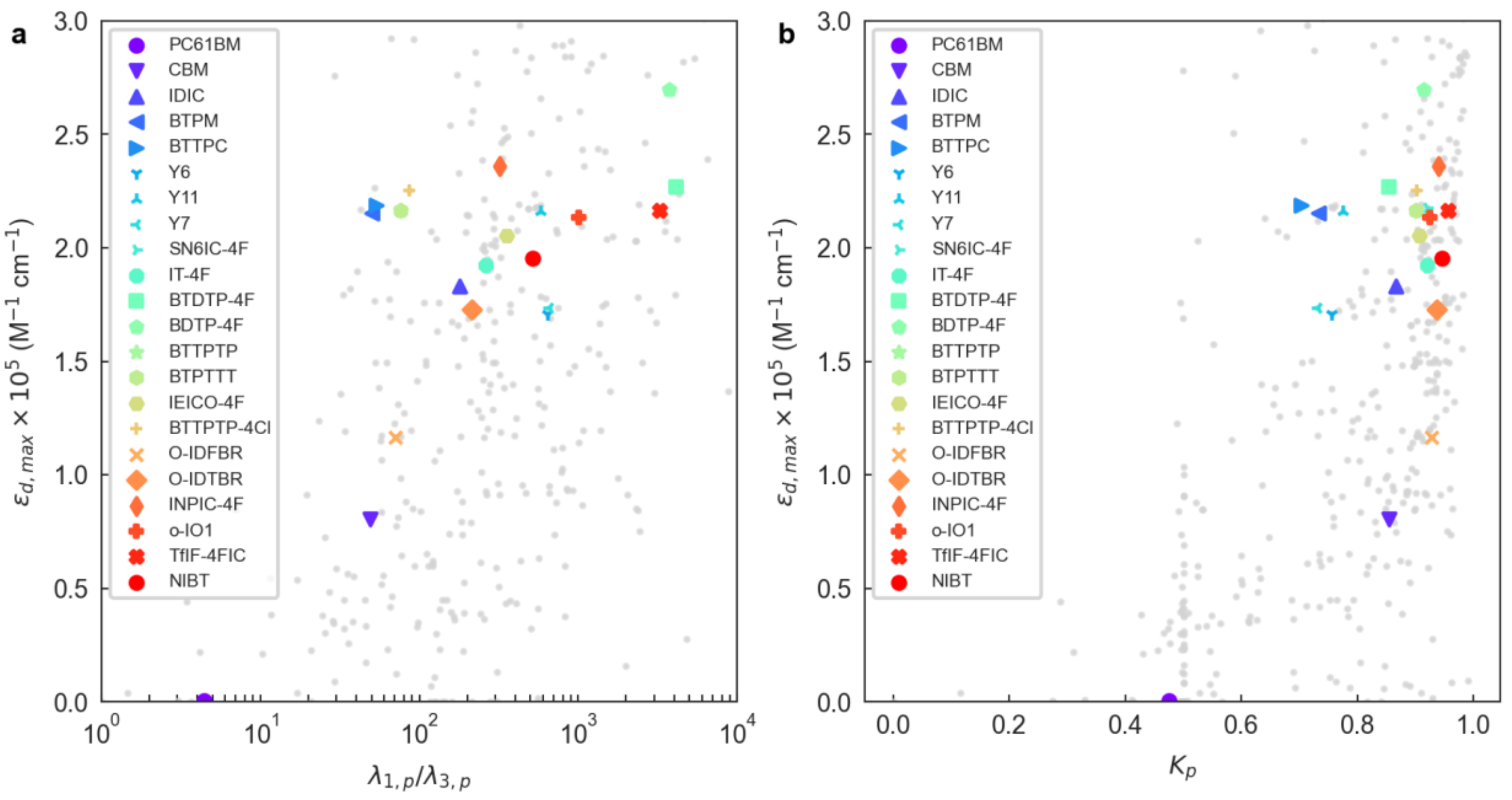

**Figure S14. Influence of the molecular planarity on the maximum molar extinction coefficient.** (a) The $\lambda_{1,p}/\lambda_{3,p}$ ratio correlates positively with $\varepsilon_{d,max}$ as straighter (more linear) molecules show larger $\lambda_{1,p}$ while enhanced molecular planarity lowers $\lambda_{3,p}$ values (which approach zero as there is no variance out of the molecular plane). (b) The shape of the molecule quantified with $K_p$ shows that linear and planar molecules (i.e., $K_p$ closer to unity)[84] enable larger $\varepsilon_{d,max}$ values. $K_p$ is defined as[84]

$$K_p = \frac{\sum_m \left| \frac{\lambda_{m,p}}{\sum_m \lambda_{m,p}} - \frac{1}{3} \right|}{4/3},$$

where $m = 1,2,3$ and $0 \leq K_p \leq 1$.

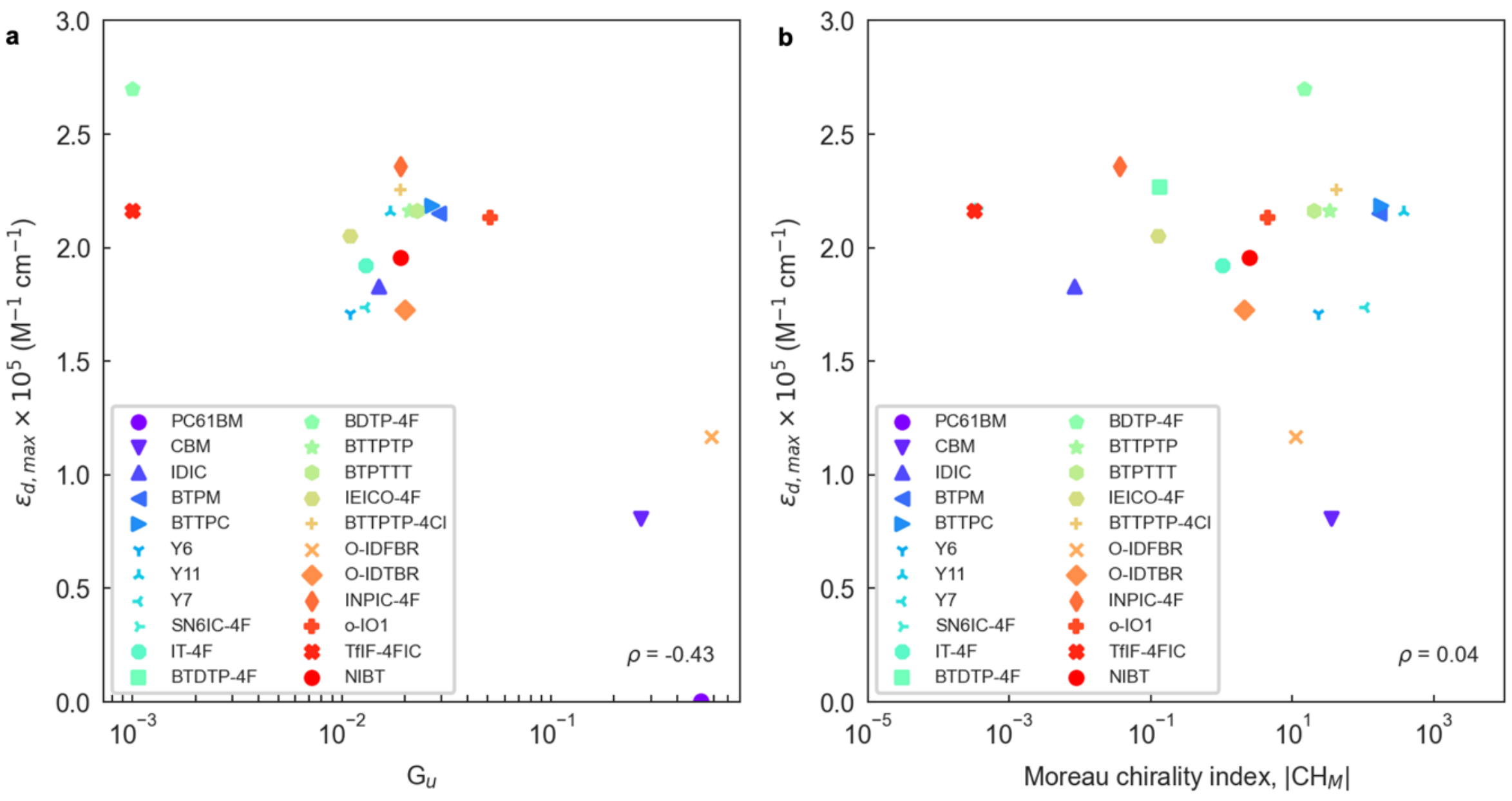

**Figure S15. Influence of molecular symmetry on absorption strength.** (a) Quantification of the total molecular symmetry as per the definition of the WHIM symmetry descriptor $G_u$ (corresponding to the unweighted geometric mean of the directional symmetries, $G_u = \sqrt[3]{\gamma_{1,u} \cdot \gamma_{2,u} \cdot \gamma_{3,u}}$)[84] shows that as the molecules lose their central symmetry (i.e., lower $G_u$ values), $\varepsilon_{d,max}$ can be further enhanced. (b) Conversely, the Moreau chirality index[111] weighted by atomic coordinates of small molecular absorbers shows poor correlation with $\varepsilon_{d,max}$.

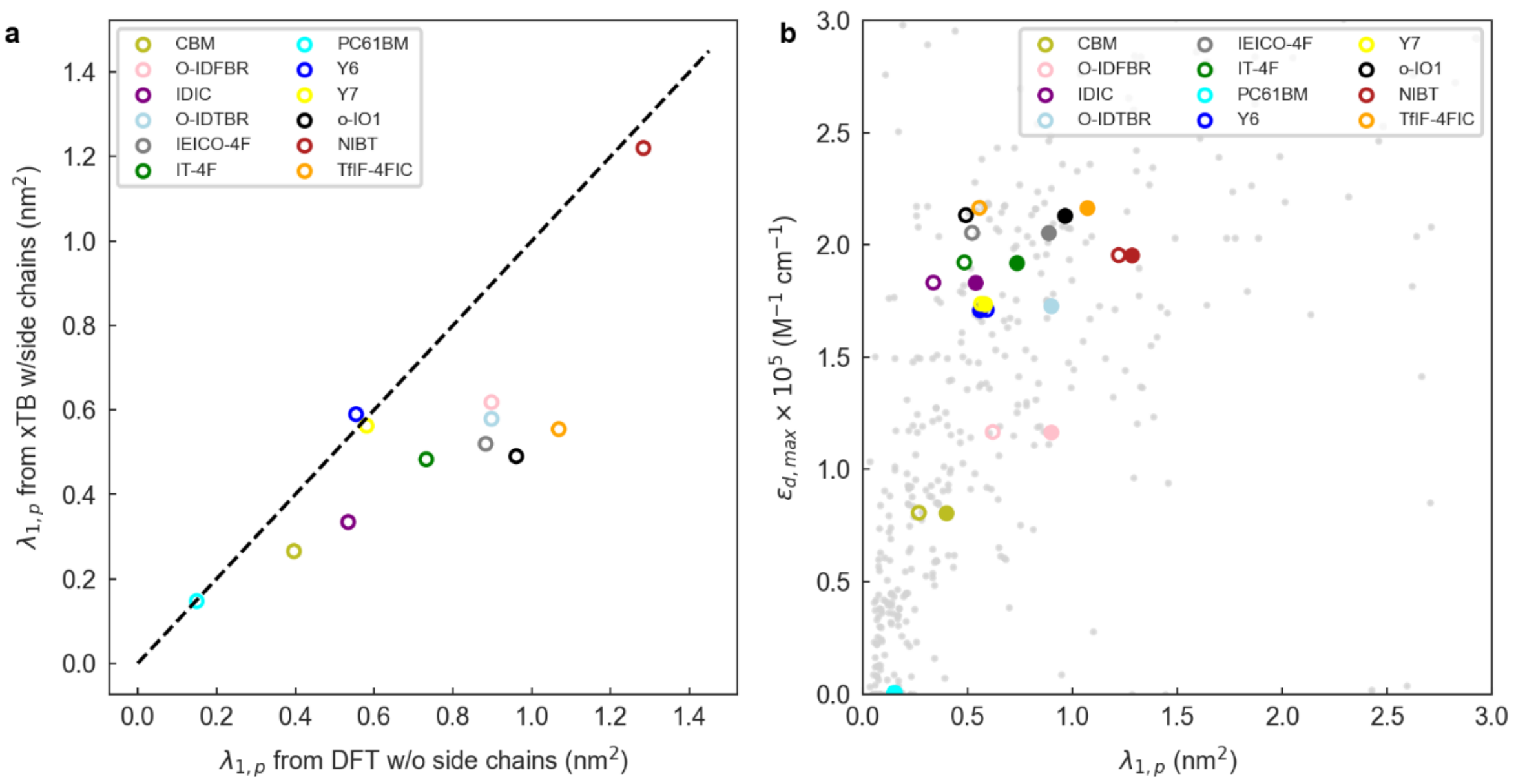

**Figure S16. Influence of the side chains on $\lambda_{1,p}$ values.** (a) Comparison of $\lambda_{1,p}$ values for a selection of small molecule acceptors as computed from xTB including side chains (y axis) and DFT-optimized geometries with methyl-substituted side chains (x axis). (b) Maximum molar extinction coefficient ($\varepsilon_{d,max}$) as a function of $\lambda_{1,p}$ for small molecule acceptors optimized with (open circles) and without (filled circles) side chains.

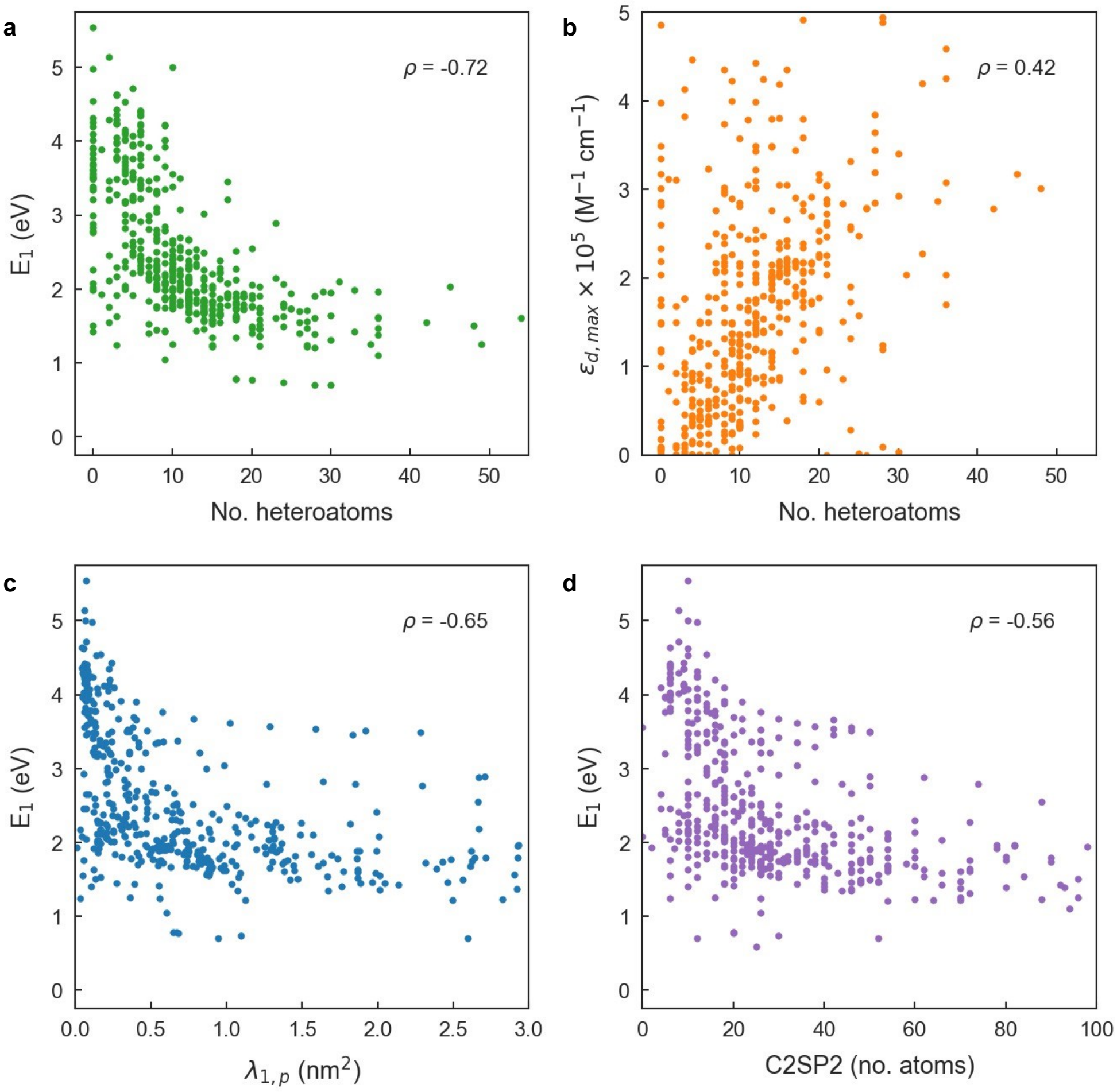

**Figure S17.** (a) Correlation between $E_1$ and the number of heteroatoms in the molecules. (b) Correlation between the molar extinction coefficient ($\varepsilon_{d,max}$) and the number of heteroatoms. (c) Correlation between $E_1$ and $\lambda_{1,p}$. (d) Correlation between $E_1$ and C2SP2. All panels include the corresponding Spearman's rank correlation coefficient ($\rho$).

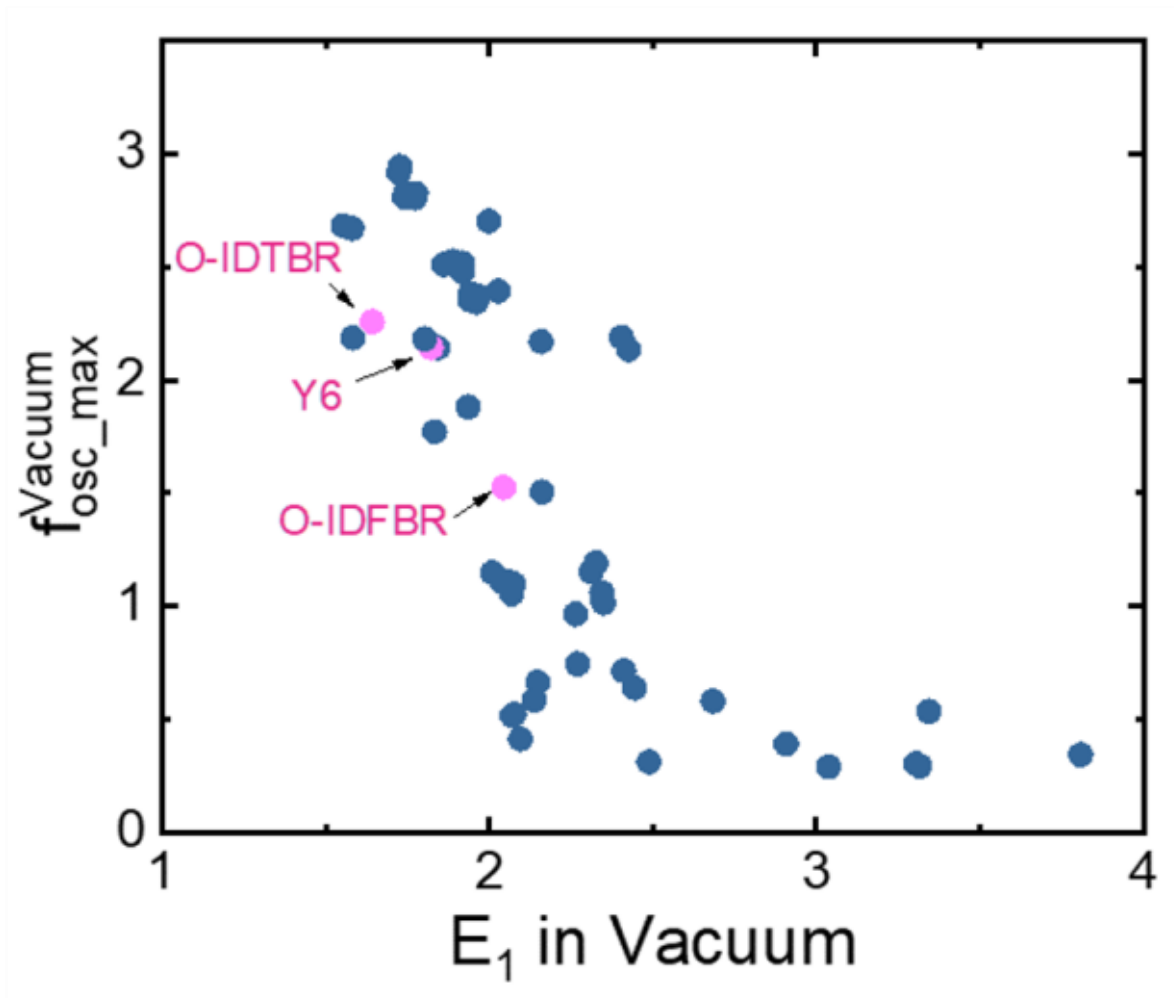

**Figure S18.** Relationship between the maximum oscillator strength and the energy of the first electronic transition in a set of TDDFT-optimized NFAs.

**Table S1.** Statistical performance of a manifold of 10-fold cross-validated baseline models using $\varepsilon_{d,max}$ as target feature.

| Model | No. variables | $R^2$ | r |
|---|---|---|---|
| 1-nearest neighbour | 2 | -0.18 ± 0.41 | 0.50 ± 0.05 |
| | 3239 | -0.10 ± 0.37 | 0.47 ± 0.09 |
| Linear regression | 2 | 0.37 ± 0.10 | 0.61 ± 0.09 |
| Random forest w/300 estimators | 2 | 0.23 ± 0.17 | 0.59 ± 0.06 |
| | 3239 | 0.65 ± 0.06 | 0.82 ± 0.03 |

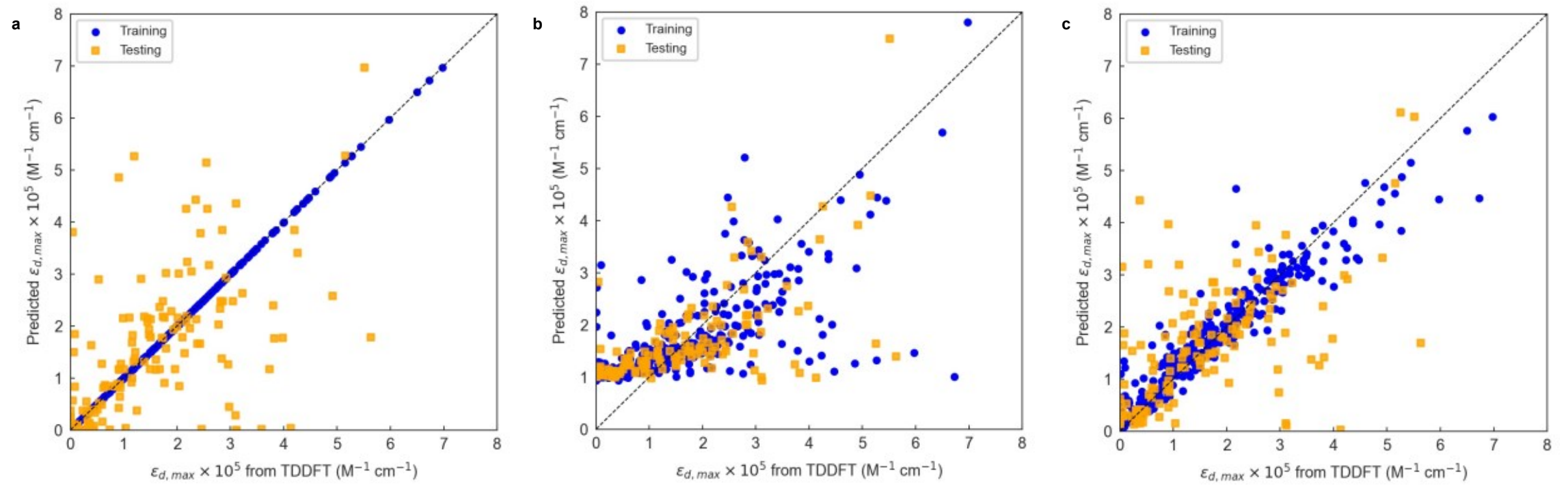

**Figure S19.** Correlation plots of three (exemplary) baseline models trained and tested on $\varepsilon_{d,max}$ using two input descriptors only: $\lambda_{1,p}$ and C2SP2. (a) 1-nearest neighbour; (b) linear regression; and (c) out-of-the-box RF trained with 300 estimators.

**Table S2.** Performance of RF models trained and 10-fold cross-validated using 300 estimators, 3 input molecular descriptors ($\lambda_{1,v}$, CIC3 and HOMO-2) and different forms of molecular fingerprint vectors. In the case of Morgan fingerprints, we set the connectivity radius to 4 units, while for topological fingerprints the minimum and maximum path counts are set to 1 and 6 units, respectively. Their vector lengths are set to either 64 or 2048 bits to reflect different degrees of model complexity.

| **No. molecular descriptors** | **Fingerprint type** | **No. bits** | **Total no. inputs** | **$R^2$** | **r** |
|---|---|---|---|---|---|
| 3 | N/A | N/A | 3 | 0.63 ± 0.06 | 0.80 ± 0.03 |
| 3 | Moiety | 159 | 162 | 0.63 ± 0.06 | 0.81 ± 0.04 |
| 3 | MACCS | 166 | 169 | 0.66 ± 0.04 | 0.83 ± 0.02 |
| 3 | Morgan | 64 | 67 | 0.70 ± 0.05 | 0.84 ± 0.03 |
| | | 2048 | 2051 | 0.69 ± 0.05 | 0.84 ± 0.03 |
| 3 | Topology | 64 | 67 | 0.68 ± 0.04 | 0.83 ± 0.02 |
| | | 2048 | 2051 | 0.69 ± 0.05 | 0.84 ± 0.03 |
| 3 | E-state | 79 | 82 | 0.66 ± 0.04 | 0.82 ± 0.02 |
| 3 | Coulomb | 320 | 323 | 0.56 ± 0.07 | 0.77 ± 0.04 |

**Table S3.** Scoring of the baseline and hyperparametrically optimized RF and ExtraTrees models, fed with 3 molecular descriptors and a Morgan fingerprint vector of 64 bits.

| Model | No. estimators | No. samples per leaf | No. samples to split | Validation | $R^2$ | r |
|---|---|---|---|---|---|---|
| RF (out-of-the-box) | 300 | 1 | 2 | 10-fold CV | 0.70 ± 0.05 | 0.84 ± 0.03 |
| RF (optimized) | 1200 | 1 | 2 | 10-fold CV | 0.70 ± 0.05 | 0.85 ± 0.03 |
| RF (optimized) | 1200 | 1 | 2 | LOOCV | 0.74 | 0.86 |
| ExtraTrees (out-of-the-box) | 300 | 1 | 2 | 10-fold CV | 0.69 ± 0.05 | 0.85 ± 0.02 |
| ExtraTrees (optimized) | 2000 | 1 | 2 | 10-fold CV | 0.70 ± 0.04 | 0.85 ± 0.02 |
| ExtraTrees (optimized) | 2000 | 1 | 2 | LOOCV | 0.73 | 0.86 |

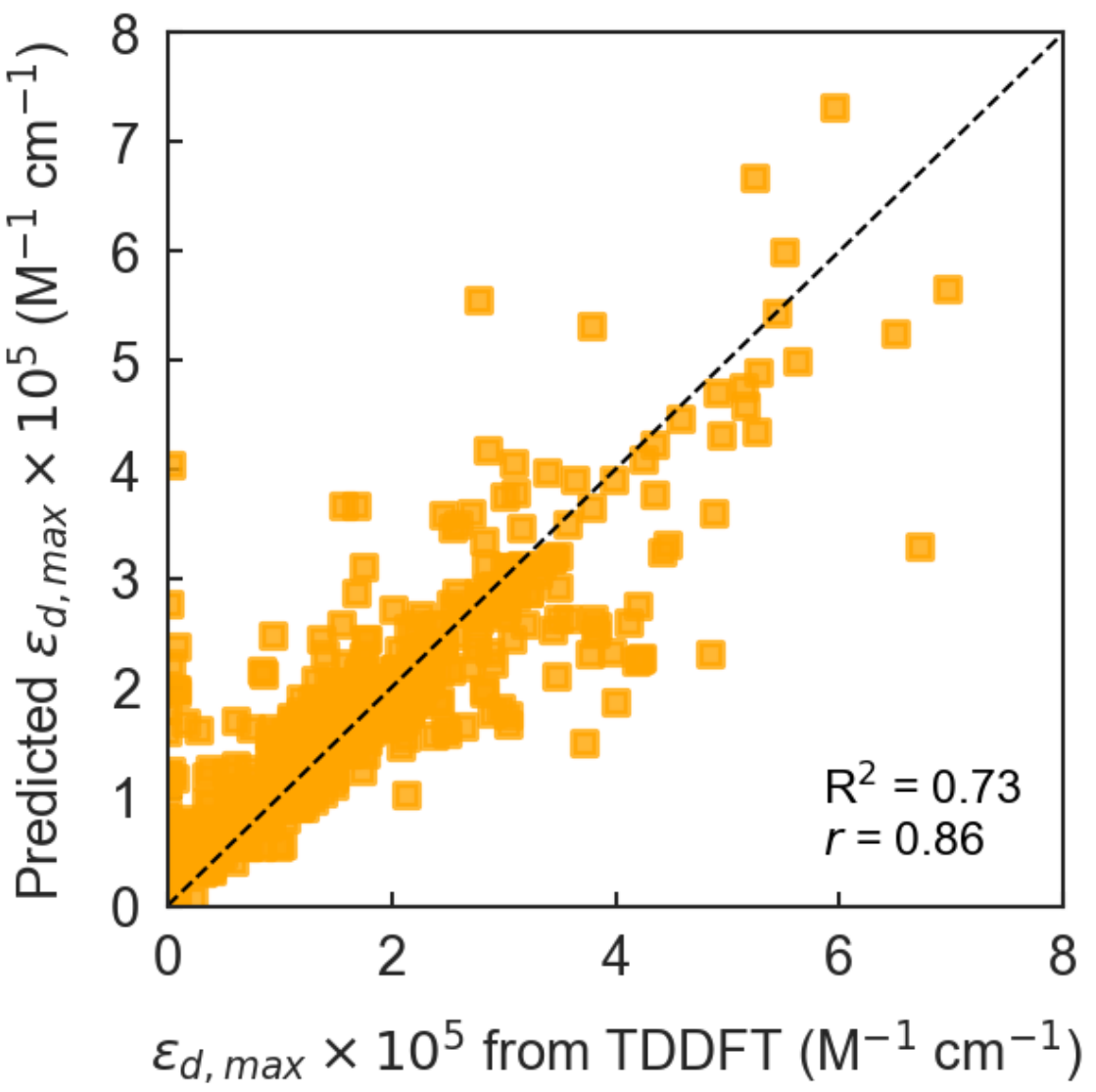

**Figure S20.** LOOCV of the optimized Extra Trees (ET) regressor fed with 3 molecular descriptors and a 64-bit vector as Morgan fingerprint.

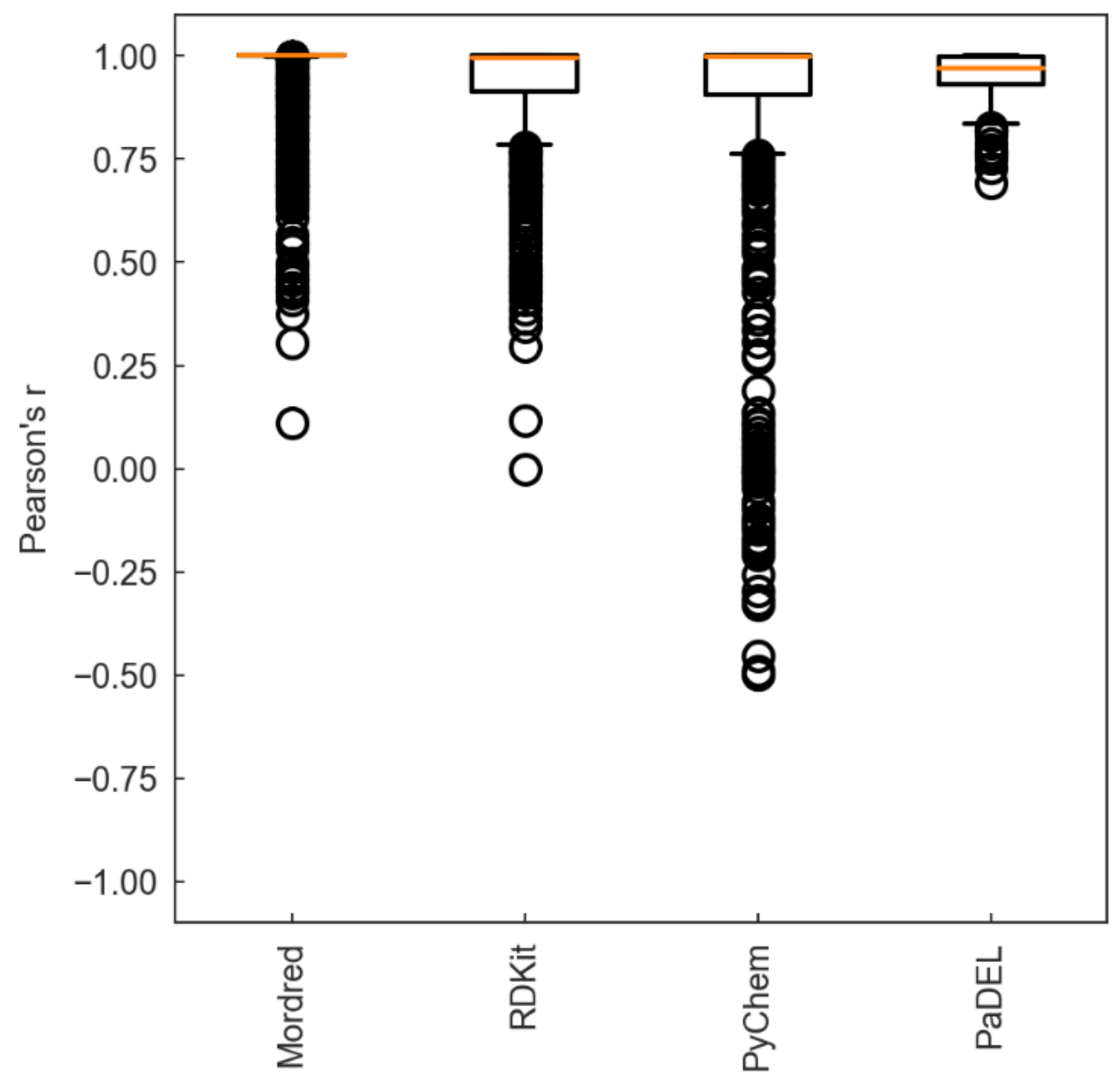

**Figure S21.** Boxplots for the Pearson correlation coefficients between different sets of molecular descriptors retrieved from xTB and DFT (B3LYP) optimized geometries.

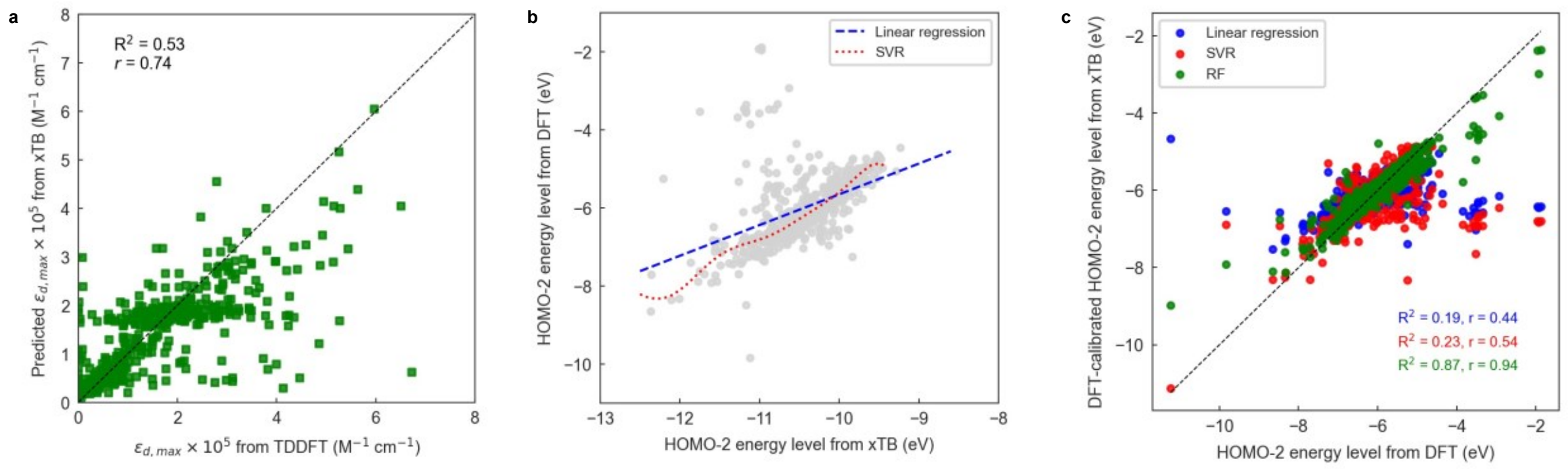

**Figure S22.** (a) Leave-one-out interpolation of a RF model trained using DFT data and tested on xTB-optimized molecules using 3 parameters ($\lambda_{1,v}$ and CIC3 recalculated from the xTB geometry, and the HOMO-2 energy level as computed in xTB) and a 64-bit vector as Morgan fingerprint. (b) Fitting of linear regression and support vector regressor (SVR) models for calibration of HOMO-2 energy levels as computed in xTB and DFT (B3LYP). The mismatch in the absolute values of HOMO-2 energy levels between DFT and xTB calculations prevents obtaining higher scorings in the RF models depicted in (a). (c) Correlation plot between HOMO-2 energy levels from DFT and the corresponding calibrated values as obtained by linear regression (blue), SVR (red) and RF (green) models. The dashed black line indicates perfect matching between DFT and calibrated values.

Supplementary Note 4. Detailed database of moieties used in this work.

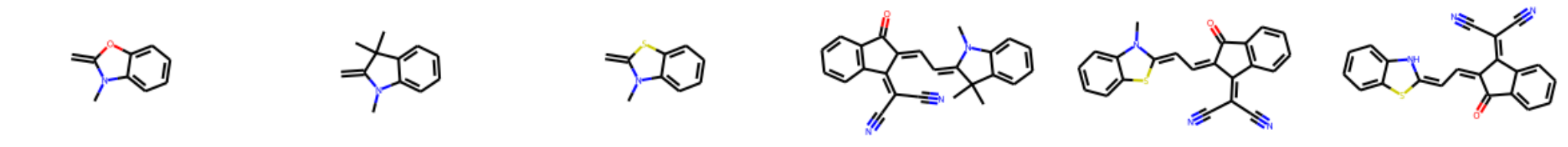

1 2 3 4 5 6

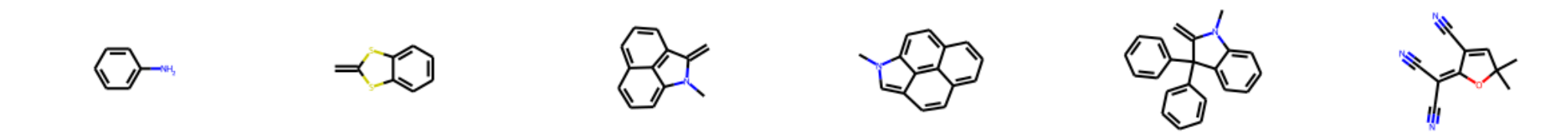

7 8 9 10 11 12

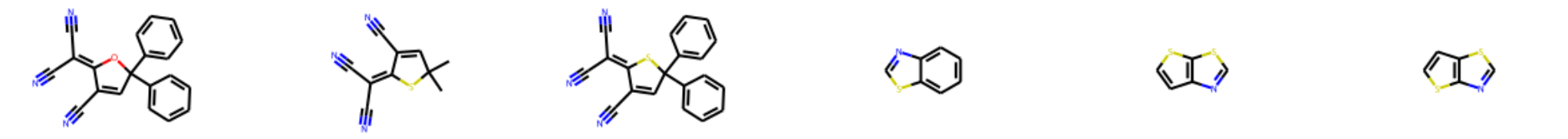

13 14 15 16 17 18

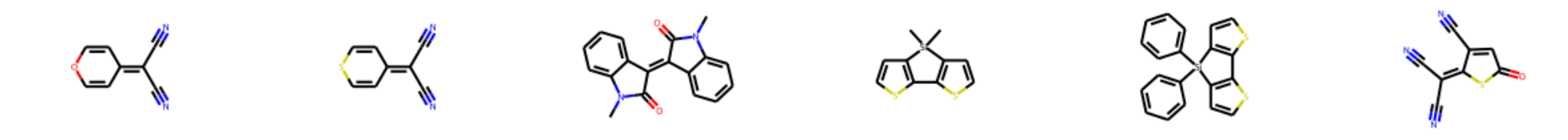

19 20 21 22 23 24

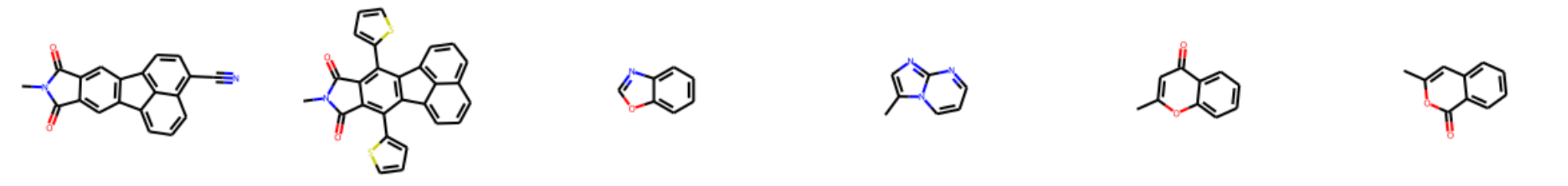

25 26 27 28 29 30

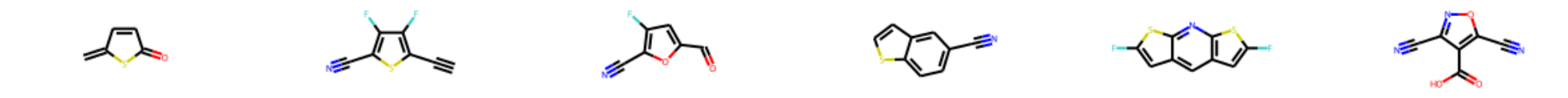

31 32 33 34 35 36

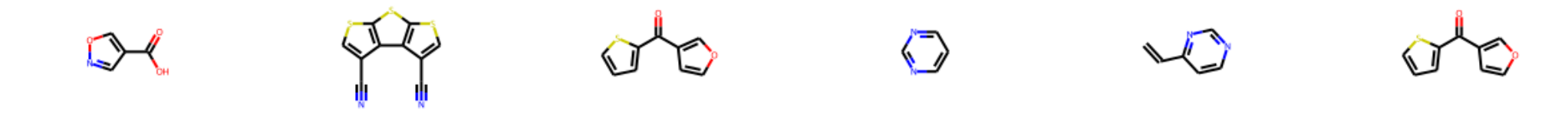

37 38 39 40 41 42

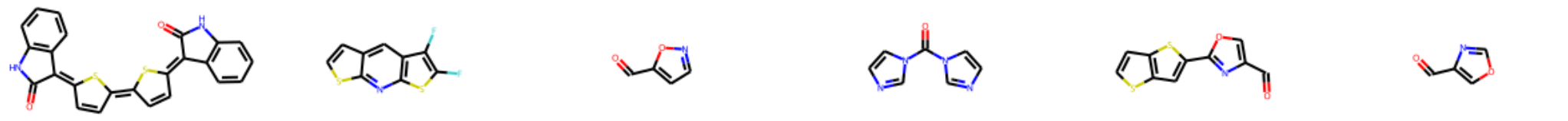

43 44 45 46 47 48

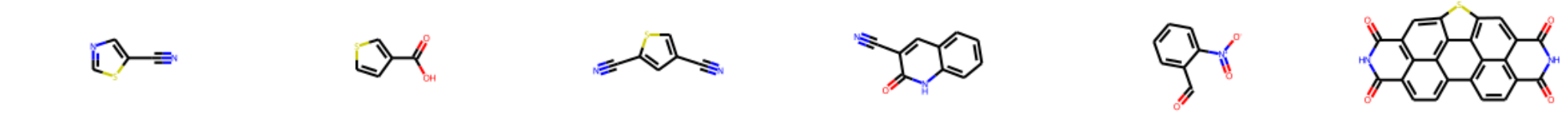

49 50 51 52 53 54

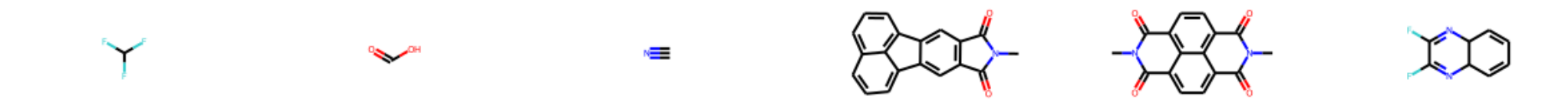

55 56 57 58 59 60

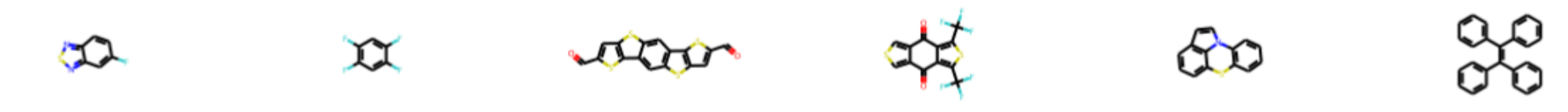

61 62 63 64 65 66

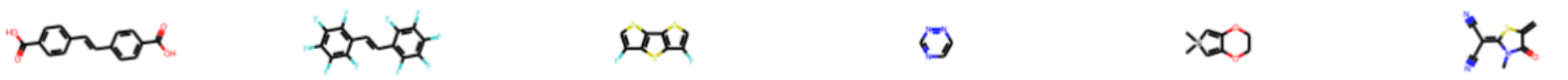

67 68 69 70 71 72

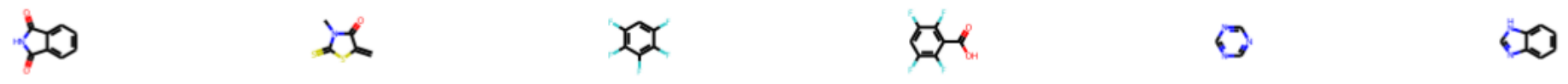

73 74 75 76 77 78

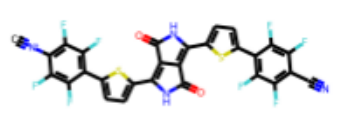

79 80 81 82 83 84

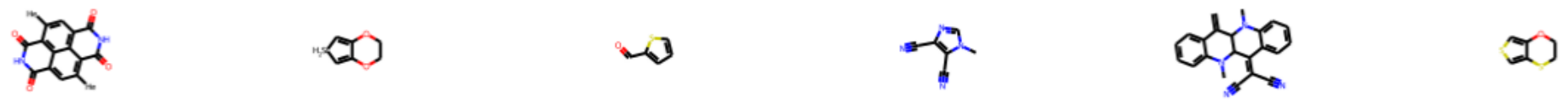

85 86 87 88 89 90

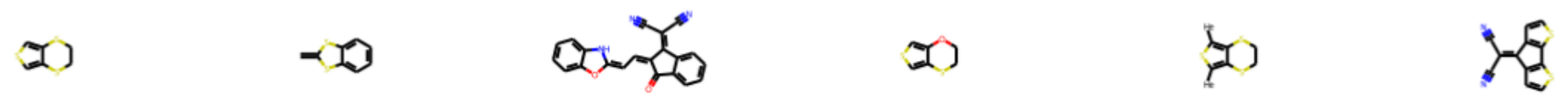

91 92 93 94 95 96

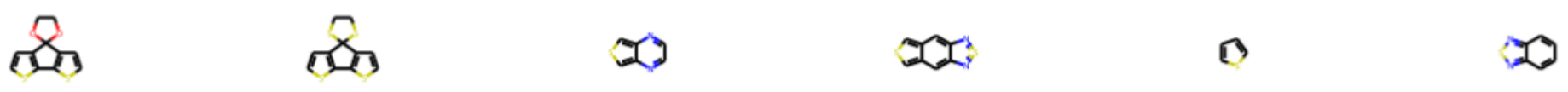

97 98 99 100 101 102

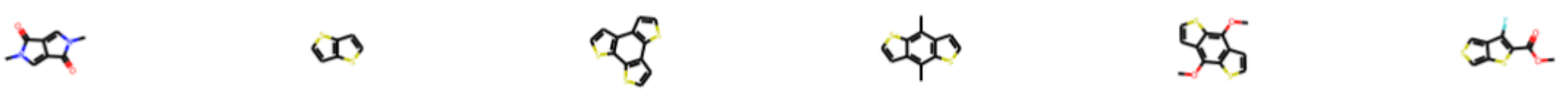

103 104 105 106 107 108

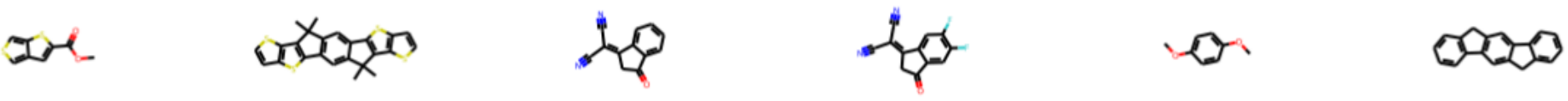

109 110 111 112 113 114

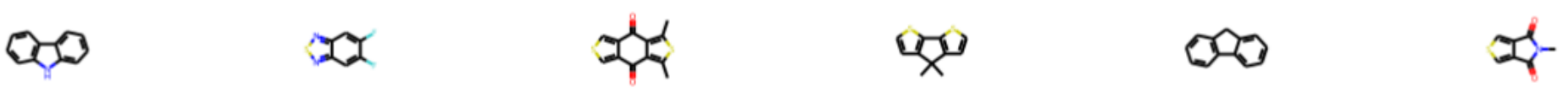

115 116 117 118 119 120

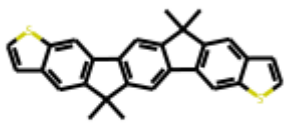
121

122
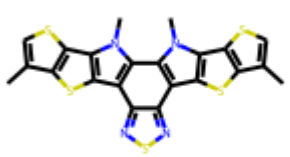
123

124
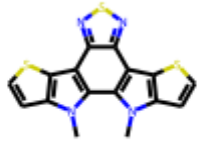
125

126

127

128

129

130

131

132

133
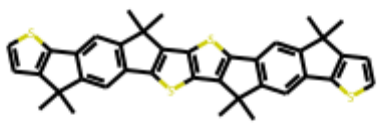
134
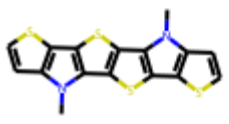
135
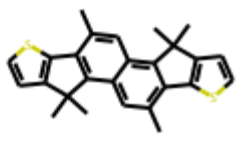
136
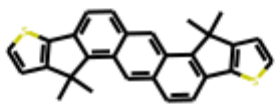
137
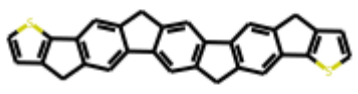
138
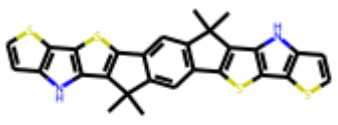
139

140

141
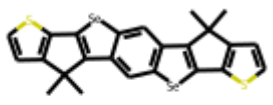
142
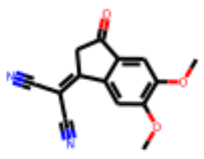
143

144

145

146

147

148
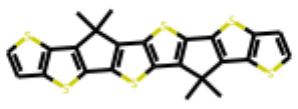
149
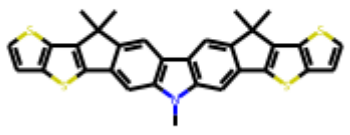
150
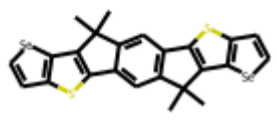
151

152

153

154

155

156

157
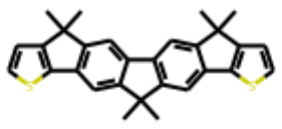
158
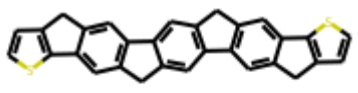
159

**Supplementary Note 5.** Estimation of computation time required to make absorption strength predictions using xTB Hamiltonians (in combination with ML models) or rigorous TDDFT.

**Table S4** provides a comparison in terms of the computation time required for the molecular geometry optimization step in the TDDFT and xTB approaches. We analysed 194 molecules from TDDFT calculations based on B3LYP/6-311+G(d,p) and 475 molecules (**Figure S6**) from xTB calculations based on GFN2–xTB to tentatively quantify the difference in computational efficiency between both methods. Our analysis suggests that geometry optimization using GFN2–xTB is ca. 3000 times faster than TDDFT/B3LYP/6-311+G(d,p), even though GFN2–xTB calculations were done on a conventional 12 CPUs laptop as opposed to the 32 CPUs dedicated cluster/workstation employed in the TDDFT calculations, thus highlighting the great advantage of using xTB over TDDFT.

Furthermore, we have estimated the time consumption for the absorption strength predictions using the established ML model in this work. The time required for the ML model training and LOOCV steps is below 10 minutes (12 CPUs), whereas the calculation of molecular descriptors (>5000 descriptors) for the full data set (479 molecules) takes no less than 180 minutes (12 CPUs). Hence, for a molecule made up of 100 atoms, the whole absorption strength determination (i.e., from geometry optimization to $\varepsilon_{d,max}$ prediction) effectively takes around 200 minutes using xTB with ML; and 1345 minutes using solely TDDFT. Nevertheless, the advantage of the ML approach is more evident as interpolation in the trained model takes less than 1 second (per molecule) to compute, which enables at least four orders of magnitude faster molecular screening with respect to TDDFT (1345 minutes or 80700 seconds per molecule).

**Table S4.** Computation time required for molecular geometry optimization steps using TDDFT/B3LYP/6-311+G(d,p) and xTB/GFN2–xTB.

| **Approach** | DFT/B3LYP/6-311+G(d,p) | xTB/GFN2–xTB |
|---|---|---|
| **No. of molecules** | 194 | 475 |
| **No. of atoms** | 18022 | 37989 |
| **No. of CPUs** | 32 | 12 |
| **Time elapsed (mins)** | 206398.355 | 136 |
| **Time elapsed per atom (mins)** | 11.45257768 | 0.003579984 |